\documentclass[aps,11pt,superscriptaddress,preprintnumbers,amsmath,amssymb,nofootinbib]{revtex4}
\usepackage{epsfig} 
\usepackage{amsmath,amsfonts} 
\usepackage{graphicx}

\large
\newcommand{\Bbar}{\,\overline{\!B}}
\newcommand{\Kbar}{\,\overline{\!K}}

\newcommand{\bbs}{\ensuremath{B_s\!-\!\Bbar{}_s\,}}
\newcommand{\bbd}{\ensuremath{B_d\!-\!\Bbar{}_d\,}}
\newcommand{\kbk}{\ensuremath{K^0\!-\!\Kbar{}^0\,}}

\newlength{\nseparation}
\def\gev{{\rm GeV}}

\def\dub{\delta_{ub}}

\def\dtpd{\delta_{t'd}}
\def\dtps{\delta_{t's}}
\def\v44{V_{4 \times 4}}
\def\v34{V_{3 \times 4}}
\def\bea{\begin{eqnarray}}
\def\eea{\end{eqnarray}}

\def\barr{\begin{eqnarray}}
\def\earr{\end{eqnarray}}

\def\ddbar{D^0-\bar D^0}

\def\la{{\lambda}}

\def\lesssim{\mathrel{\hbox{\rlap{\hbox{\lower4pt\hbox{$\sim$}}}\hbox{$<$}}}} 
\def\gtrsim{\mathrel{\hbox{\rlap{\hbox{\lower4pt\hbox{$\sim$}}}\hbox{$>$}}}}

\newcommand{\vus}{V_{us}}

\newcommand{\vcb}{V_{cb}}

\newcommand{\vub}{V_{ub}}

\newcommand{\vtpb}{V_{t'b}}

\newcommand{\vtps}{V_{t's}}
\newcommand{\vtpd}{V_{t'd}}

\def\be{\begin{equation}}
\def\ee{\end{equation}}
\def\beq{\begin{equation}}
\def\eeq{\end{equation}}

\newcommand{\bers}{\begin{eqnarray*}}
\newcommand{\eers}{\end{eqnarray*}}

\def\bbuildrel#1_#2^#3{\mathrel{\mathop{\kern 0pt#1}\limits_{#2}^{#3}}}
\def\slash#1{\setbox0=\hbox{$#1$}#1\hskip-\wd0\dimen0=5pt\advance
       \dimen0 by-\ht0\advance\dimen0 by\dp0\lower0.5\dimen0\hbox
         to\wd0{\hss\sl/\/\hss}}

\newcommand{\gae}{\lower 2pt \hbox{$\, \buildrel {\scriptstyle >}\over {\scriptstyle
\sim}\,$}}
\newcommand{\lae}{\lower 2pt \hbox{$\, \buildrel {\scriptstyle <}\over {\scriptstyle
\sim}\,$}}

\newcommand{\lt}{\left}
\newcommand{\rt}{\right}

\def\lsim{\:\raisebox{-0.5ex}{$\stackrel{\textstyle<}{\sim}$}\:}
\def\gsim{\:\raisebox{-0.5ex}{$\stackrel{\textstyle>}{\sim}$}\:}

\def\issue(#1,#2,#3){{\bf #1}, #2 (#3)}
\def\opcit(#1){ {\em op. cit.}, #1}

\def\APP(#1,#2,#3){Acta Phys.\ Polon.\ \issue(#1,#2,#3)}
\def\ARNPS(#1,#2,#3){Ann.\ Rev.\ Nucl.\ Part.\ Sci.\ \issue(#1,#2,#3)}
\def\CPC(#1,#2,#3){Comp.\ Phys.\ Comm.\ \issue(#1,#2,#3)}
\def\CIP(#1,#2,#3){Comput.\ Phys.\ \issue(#1,#2,#3)}
\def\EPJC(#1,#2,#3){Eur.\ Phys.\ J.\ C\ \issue(#1,#2,#3)}
\def\EPJD(#1,#2,#3){Eur.\ Phys.\ J. Direct\ C\ \issue(#1,#2,#3)}
\def\IEEETNS(#1,#2,#3){IEEE Trans.\ Nucl.\ Sci.\ \issue(#1,#2,#3)}
\def\IJMP(#1,#2,#3){Int.\ J.\ Mod.\ Phys. \issue(#1,#2,#3)}
\def\JHEP(#1,#2,#3){J.\ High Energy Physics \issue(#1,#2,#3)}
\def\JPG(#1,#2,#3){J.\ Phys.\ G \issue(#1,#2,#3)}
\def\MPL(#1,#2,#3){Mod.\ Phys.\ Lett.\ \issue(#1,#2,#3)}
\def\NP(#1,#2,#3){Nucl.\ Phys.\ \issue(#1,#2,#3)}
\def\NIM(#1,#2,#3){Nucl.\ Instrum.\ Meth.\ \issue(#1,#2,#3)}
\def\PL(#1,#2,#3){Phys.\ Lett.\ \issue(#1,#2,#3)}
\def\PRD(#1,#2,#3){Phys.\ Rev.\ D \issue(#1,#2,#3)}
\def\PRL(#1,#2,#3){Phys.\ Rev.\ Lett.\ \issue(#1,#2,#3)}
\def\SJNP(#1,#2,#3){Sov.\ J. Nucl.\ Phys.\ \issue(#1,#2,#3)}
\def\ZPC(#1,#2,#3){Zeit.\ Phys.\ C \issue(#1,#2,#3)}
\begin{document}


\title{Constraining the mixing matrix for Standard Model with four generations: time dependent and semi-leptonic 
CP asymmetries in $B_d^0$, $B_s$ and $D^0$}
\author{Soumitra Nandi}
\affiliation{Physique des Particules, Universit\'e de Montr\'eal,\\ C.P. 6128, succ.\ centre-ville, Montr\'eal, QC,
Canada H3C 3J7 }

\author{Amarjit Soni}
\affiliation{Physics Department, Brookhaven National Laboratory,
  Upton, NY 11973, USA}

\begin{abstract} 
Using existing experimental information from K, B and D decays as well
as electroweak precision tests and oblique parameters,
we provide constraints and correlations on the parameters of the 4X4
mixing matrix for the Standard Model with four generations
(SM4). We emphasize that some correlations amongst the parameters have important repercussions for key observables. 
We work with a particular representation of this matrix which
is highly suited for extracting information from B-decays.
Implications of the resulting constraints for time dependent and 
semileptonic CP asymmetries for $D^0$, $B^0$ and for $B_s$ are also
given.  While we show that the semi-leptonic asymmetries
may be significantly enhanced in SM4 over the SM, there are  important
constraints and correlations with other observables.
In this context we suggest that existing data from B-factories taken
on $\Upsilon (4S)$ and $\Upsilon (5S)$, and in the relevant continuum
be used to constrain the semi-leptonic asymmetries for $B_d$, $B_s$ as well as their linear
combination. Of course, the data from the Tevatron and LHCb experiments
can provide non-trivial tests of SM4 as well.
\end{abstract}
\maketitle 

\section{Introduction}
In the past few years a number of tensions in the CKM fits for the Standard Model (SM) with 3 generations have been revealed \cite{sl07,sl08,
lenz1,bona,slprl10}. There are quite serious indications that the ``predicted'' value of
$\sin 2 \beta$ is  larger compared to the value measured directly via the ``gold-plated''
$\psi K_S$ mode by as much as $\approx 3.3 \sigma$ \cite{sl10}.  Of course, the value of $\sin 2\beta$ determined from the penguin
dominated modes tends to be even smaller compared to that from the $\psi K_s$ mode
and therefore that constitutes even a larger deviation from the SM predicted value \cite{sl08}.
There are other anomalies as well that appear related. The difference in the 
partial rate asymmetries between $B^0 \to K^+ \pi^-$ and $B^+ \to K^+ \pi^0$  
is also too large \cite{pdg} to understand \cite{sl08}, though QCD complications do not allow us to draw
compelling conclusions in this regard \cite{gronau}. But with the backdrop of the hint of presence of a new CP-odd phase in the $\Delta S=1$ penguin dominated modes, it is highly suggestive that
the direct CP problem in K $\pi$ modes is receiving, at least in part, contribution
from the same new physics source. 

There are also some indications from the CDF and DO experiments at the Tevatron \cite{cdfd0}.
While the earlier indication of possible non-standard effects in $B_s \to \psi \phi$
seem to have weakened somewhat at the higher luminosity around 6/fb now being used \cite{hfag10},
D0 has announced a surprisingly large CP-asymmetry in the same sign dimuons which they attribute
primarily to originate from $B_s \to X_s \mu \nu$ \cite{d0dimuonprd,d0dimuonprl}. From  a theoretical standpoint if
new physics exists in $\Delta S=1$ B-decays, then it becomes highly unnatural for it not
to exist in $\Delta S=2$, $B_s$ mixings as well.

A simple extension of the SM with four generations (SM4) can readily account for such anomalies~\cite{SAGMN08,SAGMN10,ajb10B,buras_charm,gh10,lenz_fourth1}.
Of course, even without these anomalies, SM4 is an interesting extension of the SM worth study. The two extra phases that it possesses can give rise to a host of non-standard
CP asymmetries and in fact SM4 can significantly ameliorate the difficulties with regard to baryogenesis that SM3 has~\cite{gh08,jarlskog}.  
Besides, the heavier quarks and leptons of the 4th generation may well lead to dynamical electroweak symmetry breaking and 
thereby become useful in addressing the hierarchy problem without the need for  supersymmetry at the weak scale \cite{norton,symp_SM4_8789,Holdom:1986rn,
Hung:2009ia,Hashimoto:2009ty}. 

Motivated by these considerations we will continue our investigations of the physical implications of SM4. In particular we will    
use all the known experimental constraints such as $B \to X_s \gamma$, $B \to X_s l^+ l^-$, $\Delta M_{B_s}$, 
$\Delta M_{B_d}$, $K^+ \to \pi \nu \nu$, electroweak precision constraints from $Z \to b \bar b$ as well as oblique 
corrections~\cite{chanowitz1,chanowitz2} as in our previous work \cite{SAGMN08,SAGMN10}. However, we will now use an explicit representation of the 4X4 
CKM matrix of ~\cite{hss2_87} given long ago.  We make this particular choice as it is very well
designed to extract constraints from B decays since it was shown in a series of papers~\cite{hws87,hss1_87,hss2_87} that SM4 is highly
susceptible to those decays.

We will provide constraints and many correlations amongst the 6 real parameters and the 3 phases that enter the SM4.  
We will then apply this framework to study mixing induced CP asymmetries $S(B_d \to \psi K_s)$, $S(B_s \to \psi \phi)$, 
$S(D^0\to f)$ (where $f$ may be any self conjugate final state such as $K_s \pi^0$, 
$K_s\omega$, $K_s\rho^0$, $\pi^0\pi^0$, $K^+ K^-$, $\pi^+\pi^-$ {\it etc.}) and semi-leptonic asymmetries in $D^0$, 
$B^0$,  and in $B_s$.    

In obtaining these constraints and implications we will allow $m_{t'}$ to range from
375 to 575 GeV as suggested by current hints from study of B-decays~\cite{SAGMN08,SAGMN10}.  An interesting aspect of SM4 is that it is rather 
well constrained already. Thus, for example, while the semi-leptonic asymmetry in $B_s$ ($a_{sl}^s$) can be enhanced by as much as a factor 
of about 300 over SM3 it still cannot account for the central value of the recent D0 result \cite{d0dimuonprd}. Of course,
that observation has  only  about 2-$\sigma$ significance and therefore rather large errors
but improved experimental results could certainly rule out or confirm SM4, since
the predicted range in SM4 for  $a_{sl}^s$ is between about (0.006) to (-0.006); also its sign has to be the 
same as $S_{\psi\phi}$.  
Furthermore, for $B_d$, $a_{sl}^d$ can only be larger by around a factor of four over SM3.  
These semi-leptonic asymmetries also have interesting correlations with S($B_s \to \psi \phi$) and 
S($B_d \to \psi K_s$) respectively that should be testable.

As mentioned above one of the key difficulty for the CKM-paradigm of SM3 uncovered in recent years is that the predicted
value of $\sin 2 \beta$ is too large compared to the measured one \cite{sl07,sl10}. We will show here
that SM4 tends to alleviate this tension appreciably but at the same time then it allows to
place an important bound on $a_{sl}^d$ through the correlation mentioned in the previous para.

B-factories placed a bound on $a_{sl}^d$~\cite{hfag10} some years ago but by now they have considerable more data. 
So an improved bound would be extremely worthwhile. In the past couple of years BELLE also took substantial data on $\Upsilon$ 5S~\cite{ups5}.
In fact that data could provide a very clean study of $a_{sl}^s$ as well as 
on $A_{sl}^b$, which is defined as the linear combination of $a_{sl}^s$ and $a_{sl}^d$ \cite{d0dimuonprl}, 
since that sample provides a valuable source of this combination as well as an enriched sample of $B_s$.  CDF, D0 and LHCb should be able to 
provide very useful results on these semi-leptonic asymmetries. In fact whereas
the Tevatron  $p \bar p$ collider allowed D0 to yield the sum of $a_{sl}^d$ and $a_{sl}^s$,  the $pp$ collider 
at LHC cannot do that, but LHCb should be able to study the difference of these two asymmetries~\cite{uwer}.

We should emphasize that in this series of studies on the 4th generation \cite{SAGMN08,SAGMN10}, for simplicity,  and 
for definiteness, we have been making a tacit assumption that a heavy charge 2/3 and -1/3 quark doublet has weak interaction just like the previous three families allowing us to incorporate these readily into a 4X4 mixing-matrix resulting from an immediate generalization of the 3X3 case. Clearly if and when such a doublet of quarks is observed we will need to make detail tests on the weak interaction properties of the new quarks to verify that this assumption is correct. 

The paper is arranged as follows. After the introduction, in Sec. \ref{sec:V4G-param} and \ref{input} we provide information regarding the parametrisation 
and the constraints on the 4$\times$4 CKM matrix by incorporating oblique corrections along
with experimental data from important observables
involving Z, B and K decays as well as $B_d$ and $B_s$ mixings etc.
In Sec. \ref{results}, we present the estimates of many useful  observables in the SM4. Finally in Sec. \ref{concl}, we present our summary.

\section{Numerical Analysis}
\subsection{Parametrisation of $V_\text{CKM4}$ \label{sec:V4G-param}}
We use the parametrisation of the SM4 mixing matrix
from \cite{hss2_87}, then the elements of fourth row such as $V_{t'd}$, 
$V_{t's}$ and $V_{t'b}$, which are more relevant for the discussion  of $b$ physics, will be rather simple. Defining
\begin{align}
\vus & = \la, \qquad \vcb  = A\,\la^2, \qquad \vub = A\,\la^3\,C\,e^{-i \delta_{ub}},\nonumber 
\\
\vtps & =- Q\,\la^2\,e^{i\,\delta_{t's}}, \qquad \vtpd  = - P\,\la^3\,e^{i\,\delta_{t'd}}, 
\qquad \vtpb = - r\la
\end{align}
the generalised $4\times 4$ mixing matrix $V_\text{SM4}$ is given in eq. \ref{eqn:v4g}.
With the inputs $|V_{ub}| = (32.8 \pm 3.9)\times 10^{-4}$ and 
$|V_{cb}| = (40.86 \pm 1.0)\times 10^{-3} $ taken at $1\sigma$, constraints obtained 
on A and C are given by 
\begin{align}
 0.825\le A\le 0.865,  \hskip 40pt  0.32 \le C \le 0.42,
\end{align}
while $\la = 0.2205 \pm 0.0018$. The phase of $V_{ub}$ i.e $\delta_{ub}$ can be taken 
as the CKM angle $\gamma$ of SM3.

\begin{equation}
\mbox{\footnotesize $\left( \begin{array}{cccc}
1 - \frac{\la^2}{2}  & \la & A \la^3 C e^{-i\dub} & P \la^3 e^{-i\dtpd} \\
 & & & +Q \la^3 e^{-i\dtps}+ A C r \la^4 e^{-i\dub} \\
 & & & -P\frac{\la^5}{2}e^{-i\dtpd}\\
& & &\\
- \la  &  1 - \frac{\la^2}{2} & A \la^2 & Q \la^2 e^{-i\dtps} \\ 
 & & & + A \la^3 r - P \la^4 e^{-i\dtpd} \\
 & & & - \frac{Q}{2} \la^4 e^{-i\dtps} \\
& & & \\
A\la^3 (1 - C e^{i\dub}) & -A\la^2 & 1 - \frac{r^2 \la^2}{2} & r \la  \\
- P\,r\la^4 e^{i\dtpd} & - Q r \la^3 e^{i\dtps} & & \\
+ \frac{1}{2} A\,C \la^5 e^{i\dub}& + A \la^4 (\frac{1}{2} - C e^{i\dub}) & & \\
& & &\\
- P \la^3 e^{i\dtpd}&  - Q \la^2 e^{i\dtps}& - r \la & 1 - \frac{r^2 \la^2}{2}\\
\end{array} \right)\,. $}\label{eqn:v4g}
\end{equation}

\subsection{Inputs}\label{input}
In our earlier papers \cite{SAGMN08,SAGMN10}, to find the limits on some of the $V_{CKM4}$
elements, we concentrated mainly on the constraints that will come from
non-decoupling oblique corrections, vertex correction to $Z\to b\bar{b}$,
${\cal{BR}}(B\to X_s \gamma)$, ${\cal{BR}}(B\to X_s \, l^+ \, l^-)$,
$B_d - \bar{B_d}$ and $B_s - \bar{B_s}$ mixing,
${\cal{BR}}(K^+\to \pi^+\nu\nu)$ and the indirect {\it CP} violation in
$K_L \to \pi\pi$ described by $\epsilon_k$; we did not consider
$\epsilon'/\epsilon$ as a constraint because of the large hadronic
uncertainties, in the evaluation of its matrix elements. With the inputs given in Table \ref{tab1} we have made the
scan over the entire parameter space by a flat random number generator and
obtained the constraints on various parameters such as, $P,\, Q,\, r,\, \dtpd$ and $\dtps$ of the
4$\times$4 mixing matrix.

\begin{table}[htbp]
\begin{center}
\begin{tabular}{|c|c|}
\hline
$B_K = 0.740 \pm 0.025$ \cite{latticeold,latticenew1,latticenew} &   $R_{bb} = 0.216 \pm 0.001$ \\ 
$f_{bd}\sqrt{B_{bd}} = 0.224 \pm 0.015$ {\it GeV} \cite{Gamiz:2009ku,gamizp}& $|V_{ub}| = (32.8 \pm 2.6)\times 10^{-4}$ 
\footnote{It is the weighted average of 
$V_{ub}^{inl}=(40.1\pm 2.7 \pm 4.0)\times 10^{-4}$  and $V_{ub}^{exl}=(29.7 \pm 3.1)\times 10^{-4}$, error on the weighted average 
is increased by 50\% because of the appreciable disagreement between the two measurements.} \\
$\xi = 1.232 \pm 0.042$ \cite{Gamiz:2009ku,gamizp} & $|V_{cb}| = (40.86 \pm 1.0)\times 10^{-3} $ \\ 
$\eta_c = 1.51\pm 0.24$ \cite{uli3} & $\gamma = (73.0 \pm 13.0)^{\circ} $\\
$\eta_t = 0.5765\pm 0.0065$ \cite{buras1}& ${\cal{BR}}(B\to X_s \gamma) = (3.55 \pm 0.25)\times 
10^{-4}$\\
$\eta_{ct} = 0.494 \pm 0.046$ \cite{uli2}& ${\cal{BR}}(B\to X_s \ell^+ \ell^-) = (0.44 \pm 0.12)\times 10^{-6}$\\
$\Delta{M_s} = (17.77 \pm 0.12) ps^{-1}$  & ${\cal{BR}}(K^+\to \pi^+\nu\nu) = (0.147^{+0.130}_{-0.089})\times 10^{-9}$\\
$\Delta{M_d} = (0.507 \pm 0.005) ps^{-1}$ & ${\cal{BR}}(B\to X_c \ell \nu) = (10.61 \pm 0.17)\times 10^{-2}$\\
$|\epsilon_k|\times 10^{3} = 2.32 \pm 0.007$& $T_4 = 0.11 \pm 0.14$\\
$\kappa_{\epsilon} = 0.94 \pm 0.02$ \cite{buras_isi} \footnote{We tacitly assume that $\kappa_{\epsilon}$ in SM4 is approximately 
the same as in SM3.} & $m_t(m_t) = (163.5 \pm 1.7)$ GeV\\
\hline
\end{tabular}
\caption{Inputs that we use in order to constrain the SM4 parameter space, when not 
explicitly stated, we take the inputs from Particle Data Group \cite{pdg}; for the lattice inputs see also \cite{sl10}.}
\label{tab1}
\end{center}
\end{table}

From direct searches at the Tevatron, it follows that $m_{t'} > 335\, {\it GeV}$ \cite{cdftpm}. Taking into
account the limits from electroweak precision tests \cite{He:2001tp,Novikov:2001md,Kribs_EWPT,langacker}, 
perturbativity \cite{sher} and indications from our studies \cite{SAGMN08,SAGMN10}, plausible ranges for $m_{t'}$ and $m_{b'}$ can be taken as, 
\begin{align}
 375\,{\it GeV} < m_{t'} < 575 \,{\it GeV},\hskip 30pt  m_{b'} = (m_{t'} - 50) \, {\it GeV}.
\end{align}

Detailed mathematical formulas for the above mentioned observables (Table \ref{tab1}) can 
be seen from one of our earlier papers \cite{SAGMN10}. In this paper, we do not impose  
$S_{\psi K_s} = \sin2\beta_{eff}$ as a constraint, we show SM4 prediction for $S_{\psi K_s}$ 
and its correlation with the semileptonic asymmetry $a^d_{sl}$. 
The mathematical expression for the semileptonic asymmetry is given by
\bea
a^q_{sl} = \frac{\Delta\Gamma_q}{\Delta M_q} \tan\phi_q, \hskip 20pt (q = d, s),
\label{aslq}
\eea
where $\phi_q$ $(q = d, s)$ is defined as
\bea
\phi_q = Arg\lt[- \frac{M^s_{12}}{\Gamma^s_{12}}\rt].
\eea

Therefore the semileptonic asymmetry is the function of the CP phase $\phi_q$ and the width difference $\Delta\Gamma_q$ between the 
heavy and light mass eigenstates; $\Delta M_q$ are known with at least 1\% accuracy \cite{pdg}. SM prediction for $\Delta\Gamma_q$ has an 
overall impact on the results for semileptonic asymmetry.
At leading order in $\alpha_s$ we do not have appreciable SM4 contribution to $\Delta\Gamma_q$; we use the SM predictions  
for $\Delta\Gamma_q$ \cite{uli_lenz} to find out the allowed ranges for the semileptonic asymmetries.   

In addition, we study $D^0-\bar D^0$ mixing in the 
presence of a fourth generation of quarks. In particular, we calculate the size of the 
allowed CP violation, which could be large compared to the SM, and show its 
parametric dependence on CKM4 elements.

Within the SM, $D^0-\bar D^0$ mixing proceeds to an excellent approximation only
 through the box diagrams with internal $b$ and $s$ quark exchanges. In the case of
four generations there is an additional contribution to  $D^0-\bar D^0$ mixing
coming from the virtual exchange of the fourth generation down quark $b'$.

The short distance (SD) contributions to the matrix element of the $\Delta C=2$ effective 
Hamiltonian can be written as
\begin{align}
 \langle \bar D^0 | {\mathcal H}_{\rm eff}^{\Delta C = 2} | D^0\rangle_{\rm SD} &\,\equiv\, \left| M_{12}^D\right| e^{2i\phi_D}\,=\, \left(M_{12}^D\right)^\ast\,,
\end{align}
where
\begin{align}
 M_{12}^D \,&=\, \frac{G_F^2}{12\pi^2}f_D^2 \hat B_D m_D M_W^2 \overline{M}_{12}^D\,,
\end{align}
with
\begin{align}\label{M12D}
\begin{split}
\overline M_{12}^D \,&=\,{\lambda_{s}^{(D)}}^{\ast 2}\eta^{(K)}_{cc}S_0(x_s)+{\lambda_{b}^{(D)}}^{\ast 2}\eta^{(K)}_{cc}S_0(x_b)+{\lambda_{b^\prime}^{(D)}}^{\ast 2}\eta^{(K)}_{tt}S_0(x_{b^\prime})\\
+&\,2{\lambda_{b}^{(D)}}^\ast {\lambda_{s}^{(D)}}^\ast \eta^{(K)}_{cc}S_0(x_b,x_s)+2{\lambda_{b^\prime}^{(D)}}^\ast {\lambda_{s}^{(D)}}^\ast \eta^{(K)}_{ct}S_0(x_{b^\prime},x_s)+2{\lambda_{b^\prime}^{(D)}}^\ast {\lambda_{b}^{(D)}}^\ast \eta^{(K)}_{ct}S_0(x_{b^\prime},x_b)\,,
\end{split}
\end{align}
where 
\begin{align}
 {\lambda_{i}^{(D)}} &= V_{ci}^\ast V_{ui}\,\,\,(i=s,b,b').
\end{align}
For the QCD corrections we will use the approximate relations
\begin{align}\label{eq:QCDcorr}
 \eta^{(D)}_{b^\prime b^\prime} &\approx \eta^{(K)}_{tt}\,, & \eta^{(D)}_{b^\prime b}\approx \eta^{(D)}_{b^\prime s}&\approx \eta^{(K)}_{ct}\,, &
\eta^{(D)}_{ss}\approx\eta^{(D)}_{bb}\approx\eta^{(D)}_{bs}&\approx\eta^{(K)}_{cc}\,.
\end{align}

Including the long distance part the full matrix elements are given by,
\begin{align}
\langle \bar D^0 | {\mathcal H}_{\rm eff}^{\Delta C = 2} | D^0\rangle &\,=\, \left( M_{12}^D+M_{12}^{\rm LD}\right)^\ast - \frac{i}{2}{\Gamma_{12}^{\rm LD}}^\ast\,,\\
\langle D^0 | {\mathcal H}_{\rm eff}^{\Delta C = 2} | \bar D^0\rangle &\,=\, \left( M_{12}^D+M_{12}^{\rm LD}\right) - \frac{i}{2}{\Gamma_{12}^{\rm LD}}\,.
\end{align}
Here $\Gamma_{12}^{\rm LD}$ and $M_{12}^{\rm LD}$ stand for long distance (LD) contributions 
with the former arising exclusively from SM3 dynamics. These contributions are very 
difficult to estimate reliably; we scan flatly over the intervals 
\cite{Bigi:2009df,Altmannshofer:2010ad}. 
\begin{align}
-0.02\,{\rm ps}^{-1}\leq & M_{12}^{\rm LD} \leq 0.02\,{\rm ps}^{-1}\,,\\
-0.04\,{\rm ps}^{-1}\leq & \Gamma_{12}^{\rm LD}\leq 0.04\,{\rm ps}^{-1}\,.
\end{align}

$\ddbar$ oscillations can be characterised by the normalised mass and width differences 
\begin{align}
x_D &\equiv \frac{\Delta M_D}{\bar \Gamma} \,, &y_D &\equiv \frac{\Delta\Gamma_D}{2\bar\Gamma}\,, &\bar\Gamma &= \frac{1}{2}(\Gamma_1 + \Gamma_2),
\end{align}
with 
\begin{align}
\Delta M_D &= M_1 - M_2 = 2 {\it Re}\left[\frac{q}{p}(M^D_{12}-\frac{i}{2}\Gamma^D_{12})\right]
\nonumber\\
 & = 2 {\it Re}\sqrt{|M^D_{12}|^2-\frac{1}{4}|\Gamma^D_{12}|^2 - i 
Re(\Gamma^D_{12} {M^D_{12}}^{\ast})},
\end{align}
\begin{align}
\Delta\Gamma_D &= \Gamma_1 - \Gamma_2 = -4 {\it Im}\left[\frac{q}{p}(M^D_{12}-\frac{i}{2}
\Gamma^D_{12})\right]
\nonumber\\
 & = -4 {\it Im}\sqrt{|M^D_{12}|^2-\frac{1}{4}|\Gamma^D_{12}|^2 - i Re(\Gamma^D_{12} {M^D_{12}}^{\ast})},
\end{align}
where 
\begin{align}
\frac{q}{p} &\equiv \sqrt{\frac{{M^D_{12}}^{\ast} - \frac{i}{2}{\Gamma^D_{12}}^{\ast}}
{M^D_{12} - \frac{i}{2}\Gamma^D_{12}}}.
\label{eq:qp}
\end{align}

For practical purposes it is sufficient to consider the time-dependent CP asymmetry 
$S_f$ as \cite{Bigi:2009df}
\begin{equation}
\frac{\Gamma(D^0(t) \to f) - \Gamma(\bar D^0(t) \to f)}
{\Gamma (D^0(t) \to f) + \Gamma(\bar D^0(t) \to f)}
\equiv S_{f}(D) \frac{t}{ 2\overline\tau _D} \,,
\label{eq:GASYM}
\end{equation}
which is given by
\begin{equation}
 \eta_f  S_f(D)\simeq - \left[ y_D\left(\left| \frac{q}{p}\right| -\left| \frac{p}{q}\right|   \right)\cos\varphi -
x_D \left(\left| \frac{q}{p}\right| +\left| \frac{p}{q}\right|   \right) \sin\varphi \right] \,,
\label{eq:Sf}
\end{equation}
where $\eta_f=\pm 1$ is the CP parity of the final state $f$.
The SM3 prediction for $\eta_fS_f(D)$ is
\begin{align}
\left[\eta_f S_f(D)\right]_\text{SM3}&\approx\,-2 \cdot 10^{-6}\,.
\end{align}

Finally, the semileptonic asymmetry is defined as
\begin{equation}\label{eq:ASL}
a_\text{SL}(D) \equiv  \frac{\Gamma (D^0(t) \to \ell^-\bar\nu K^{+(*)}) - \Gamma (\bar D^0 \to \ell^+\nu K^{-(*)})}
{\Gamma (D^0(t) \to \ell^-\bar\nu K^{+(*)}) + \Gamma (\bar D^0 \to \ell^+\nu K^{-(*)})}=
\frac{|q|^4 - |p|^4}{|q|^4 + |p|^4}
\approx 2\left(\left|\frac{q}{p}\right|-1\right)\,.
\end{equation}

The world averages based on data from BaBar, Belle and CDF are given by 
\cite{Barberio:2007cr,Barberio:2008fa,Schwartz:2009jv}
\begin{align}\label{eq:expdata}
x_D  &= \left(0.98^{+0.24}_{-0.26}\right)\%\,, &y_D  &= \left(0.83\pm 0.16\right)\%\,,\nonumber\\
|q/p| &=\left(0.87^{+0.17}_{-0.15}\right)\,, &\phi &=\left(-8.5^{+7.4}_{-7.0}\right)^\circ\,,\\
\eta_f S_f(D) &=\,\left(-0.248\pm 0.496\right)\%\,,\nonumber
\end{align}
with $\phi$ being the phase of $q/p$ and the asymmetry $\eta_fS_f(D)$ defined in (\ref{eq:GASYM}).

In addition to Table \ref{tab1} the relevant input parameters for $D^0-\bar D^0$ mixing 
are given in Table \ref{tab:parameters}.  
\begin{table}[t]
\begin{center}
\begin{tabular}{|l|l||l|l|}
\hline
parameter & value & parameter & value \\
\hline\hline
$m_{D}$ & $(1.86484\pm0.00017)$GeV & $\bar\tau_{D}$ & $(0.4101\pm 0.0015)$ps\\
$f_D$& $(0.212\pm {0.014})\gev$\hspace{12pt}\cite{Lubicz:2008am}& $m_c(m_c)$ & $(1.268\pm 0.009) \gev$\hspace{12pt}\cite{Laiho:2009eu,Allison:2008xk}  \\
$\hat B_D$& $1.18^{+0.07}_{-0.05}$ \cite{buras_charm,soni_yamada} & $m_b(m_b)$ & $(4.20^{+0.17}_{-0.07}) \gev$\hfill\cite{Barberio:2007cr}  \\
\hline
\end{tabular}
\caption{Values of the input parameters for $D$ mesons used in our analysis.} \label{tab:parameters}
\end{center}
\end{table}

\subsection{Results}\label{results}

\begin{table}[htbp]
\begin{center}
\begin{tabular}{|l|l||l|l|}
\hline
parameter & allowed range & parameter & allowed range \\
\hline\hline
$\la$ & $0.2205 \pm 0.0018$ & $|V_{t'b}|$ & $<$ 0.12 \\
C & $0.32 \to 0.42$ &  $|V_{t'd}|$ & $<$ 0.05 \\
A & $0.825 \to 0.865$ & $|V_{t's}|$ & $<$ 0.11 \\
$\gamma$ & $(73\pm 13)^{\circ}$ & $|V_{ub'}|$ & $<$ 0.05 \\
r & $<$ 0.5 & $|V_{cb'}|$ & $<$ 0.11  \\
P & $<$ 5.0 & $|\la^{t'}_{db}|$ & $<$ 0.002\\ 
Q & $<$ 2.5 & $|\la^{t'}_{sb}|$ & $<$ 0.01\\ 
  & & $|\la^{b'}_{uc}|$ & $<$ 0.0025 \\
\hline
\end{tabular}
\caption{Allowed ranges of the CKM4 parameters obtained from our analysis.} \label{tab:fitr}
\end{center}
\end{table}

\begin{figure}[t]
\includegraphics[width=.48\textwidth]{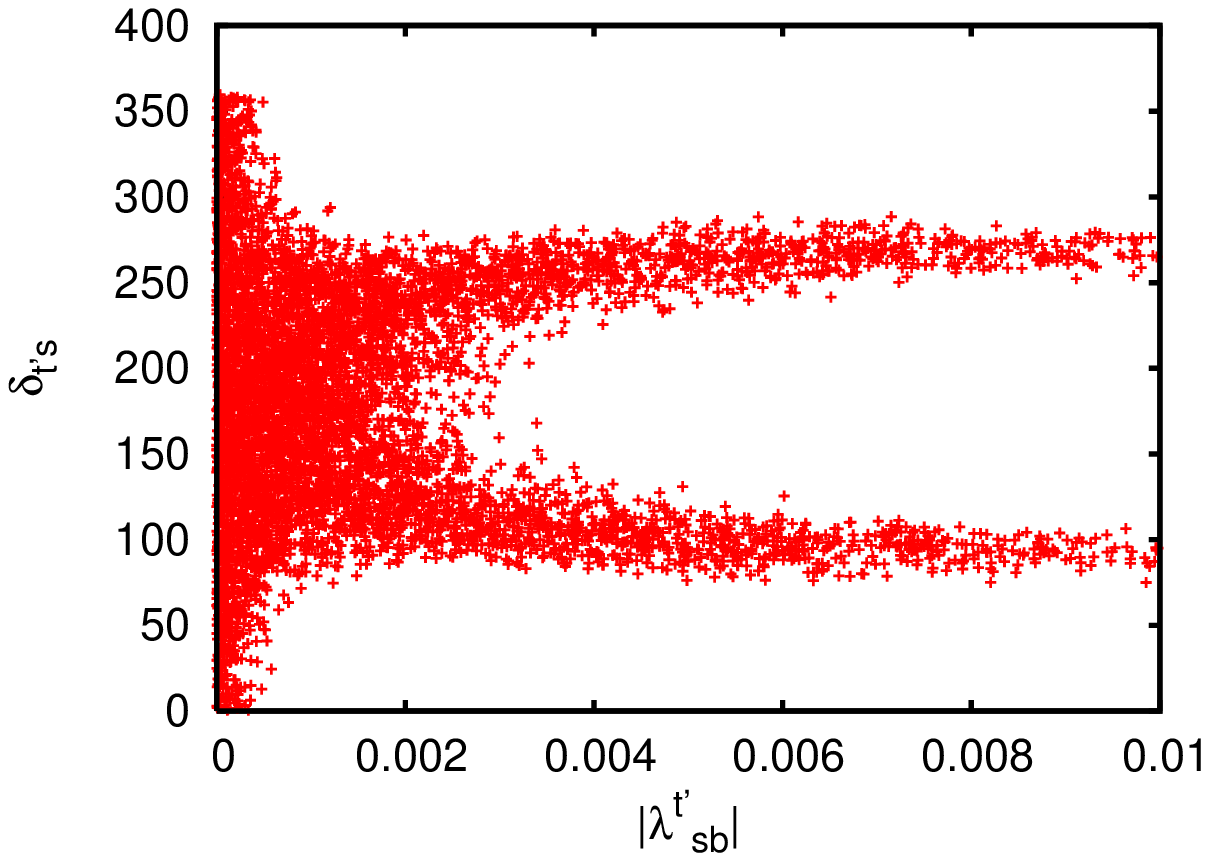}\hspace{.03\textwidth}
\includegraphics[width=.48\textwidth]{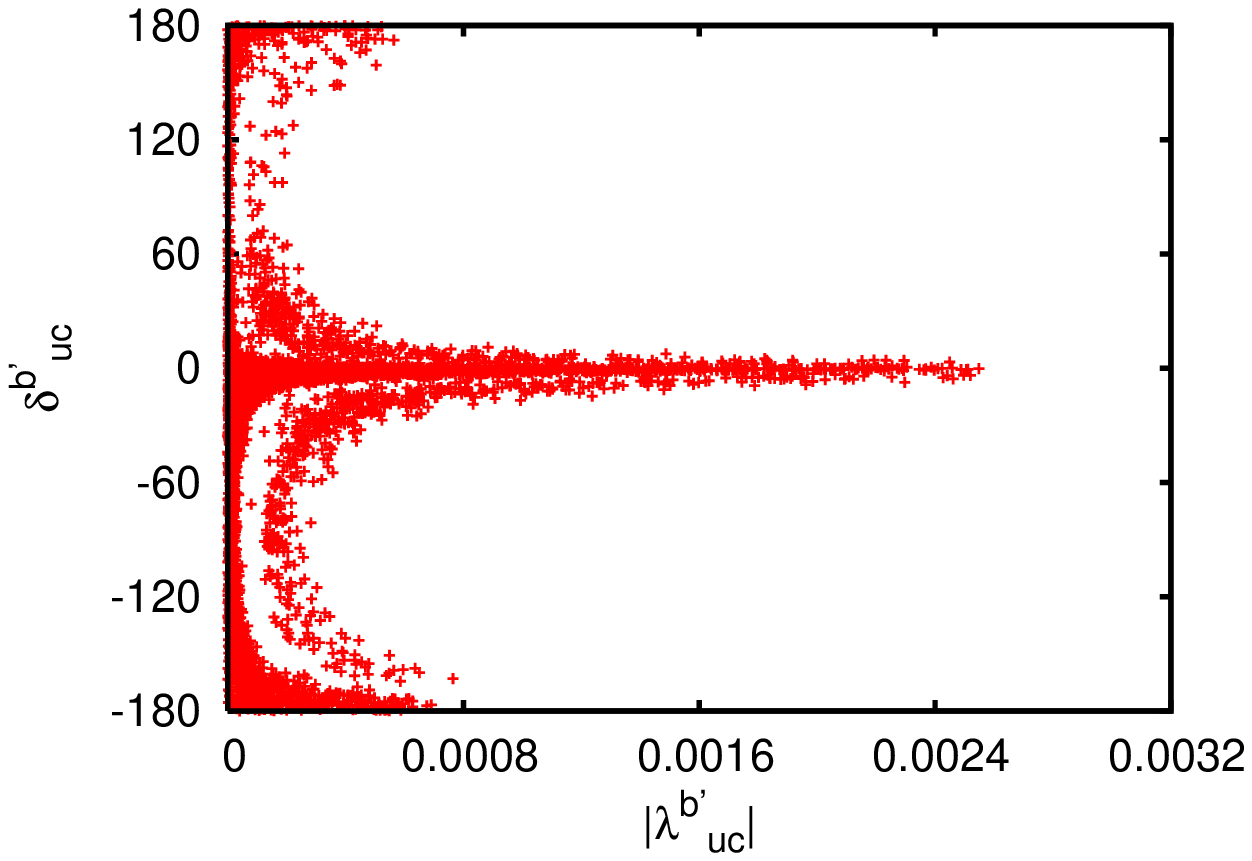}
\caption{Fourth generation parameter space; left panel shows the variation of 
$|\la^{t'}_{sb}| = |V^{\ast}_{t's} V_{t'b} |$ with the phase $\delta_{t's}$ of $V_{t's}$,
whereas right panel shows it for $|\la^{b'}_{uc}| = |V^{\ast}_{ub'} V_{cb'} |$ with 
$\delta^{b'}_{uc}$, the phase difference between the phase of $V^{\ast}_{ub'}$ and $V_{cb'}$.}
\label{fig1}
\end{figure}

Allowed ranges for different CKM4 parameters/elements are given in Table \ref{tab:fitr}.
Constraint on $V_{t'b}$ or equivalently on the new parameter $r$ (i.e $V_{t'b} = -\,r \la$) is obtained from
non-decoupling oblique corrections ($T_4$) and vertex corrections to $Z\to b\bar{b}$.
We also note the allowed ranges for the product of the different 
CKM4 elements, $|\la^{t'}_{db}| = |V^{\ast}_{t'd} V_{t'b} |$, $|\la^{t'}_{sb}| = |V^{\ast}_{t's} V_{t'b} |$ 
and $|\la^{b'}_{uc}| = |V^{\ast}_{ub'} V_{cb'} |$, obtained from our analysis; these are relevant to $B_d^0$, $B_s$ and $D^0$ oscillations. 
Allowed ranges for the corresponding 
phases and their correlations with the magnitude of the product couplings are shown in Fig. \ref{fig1}.
We note that values of $|\la^{t'}_{sb}|$ larger than $0.002$ correspond to very narrow regions of the phase 
$\delta_{t's}$ (left panel, Fig. \ref{fig1}) close to $90^{\circ}$ or $270^{\circ}$, whereas that 
for $\delta^{b'}_{uc}= \delta_{cb'} - \delta_{ub'}$ (right panel) is close to zero when $|\la^{b'}_{uc}| \gsim 0.0008$. 
$\bbd$ and $\bbs$ mixing are sensitive to the new 
parameters $[P,\dtpd]$ and $[Q,\dtps]$ respectively whereas $\kbk$ and $\ddbar$ mixing are sensitive to all these
four new parameters and their parametric dependencies are given by,

\begin{align}
\la^{t'}_{ds} &= V^{\ast}_{t's} V_{t'd} = P\,Q\,\la^5\,e^{i(\dtpd-\dtps)}\nonumber\\
\la^{b'}_{uc} &= V^{\ast}_{ub'} V_{cb'} =Q\,\la^5\,\left(Q + P e^{i(\dtpd-\dtps)}\right)
\label{paradep}
\end{align} 
respectively. In this framework it is quite natural to expect that there is a strong 
correlation between $\kbk$ and $\ddbar$ mixing, as pointed out in the case of purely 
left-handed currents \cite{nir_grossman, Bigi:2009df}, $\ddbar$ mixing is also correlated 
with the observables from $B_d$ and $B_s$ mixing and decays. 
So the constraints obtained on the new parameters from the inputs given in Table \ref{tab1}
, especially $\epsilon_K$ and $Br(K^+\to \pi^+\nu\bar\nu)$, are helpful to find the allowed 
parameter space for $\la^{b'}_{uc}$ and the corresponding phase difference $\delta^{b'}_{uc}$.

\begin{figure}[t] 
\includegraphics[width=.48\textwidth]{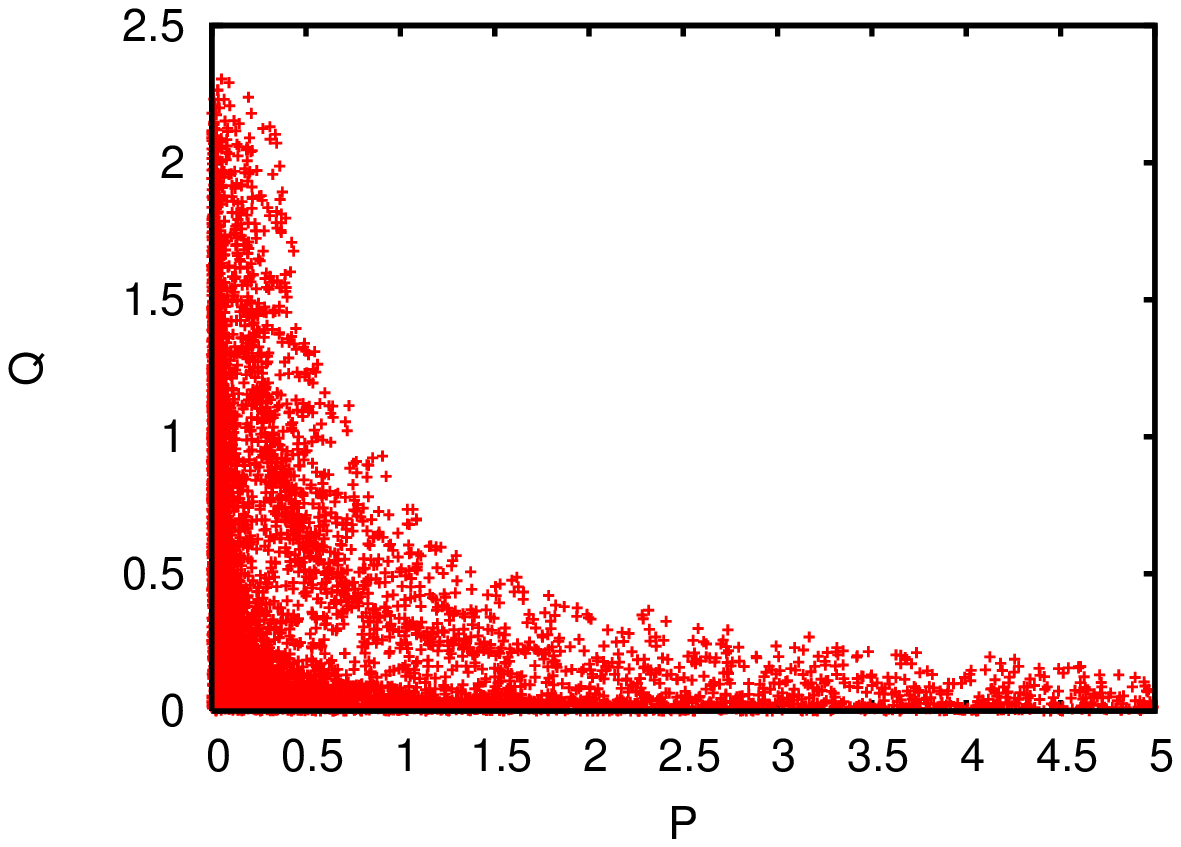}\hspace{.03\textwidth}
\includegraphics[width=.48\textwidth]{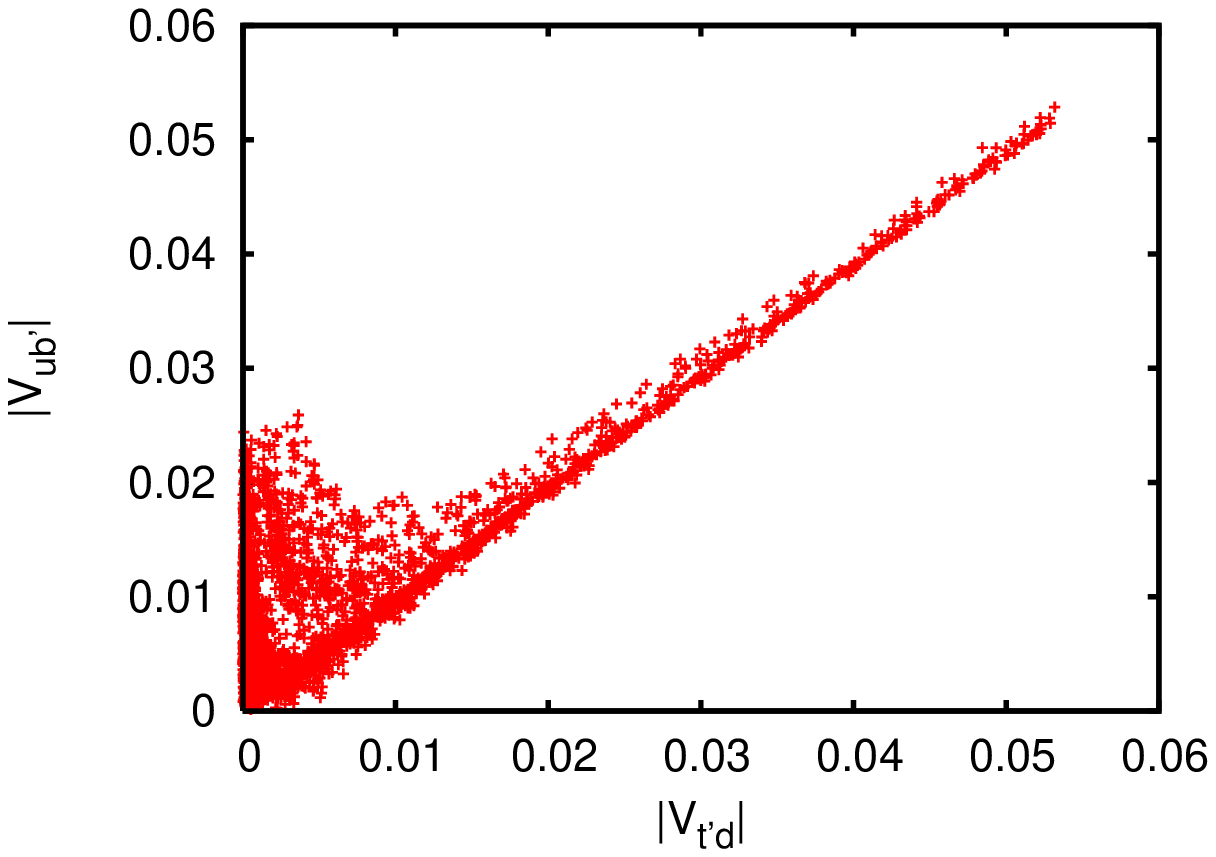}\\
\includegraphics[width=.48\textwidth]{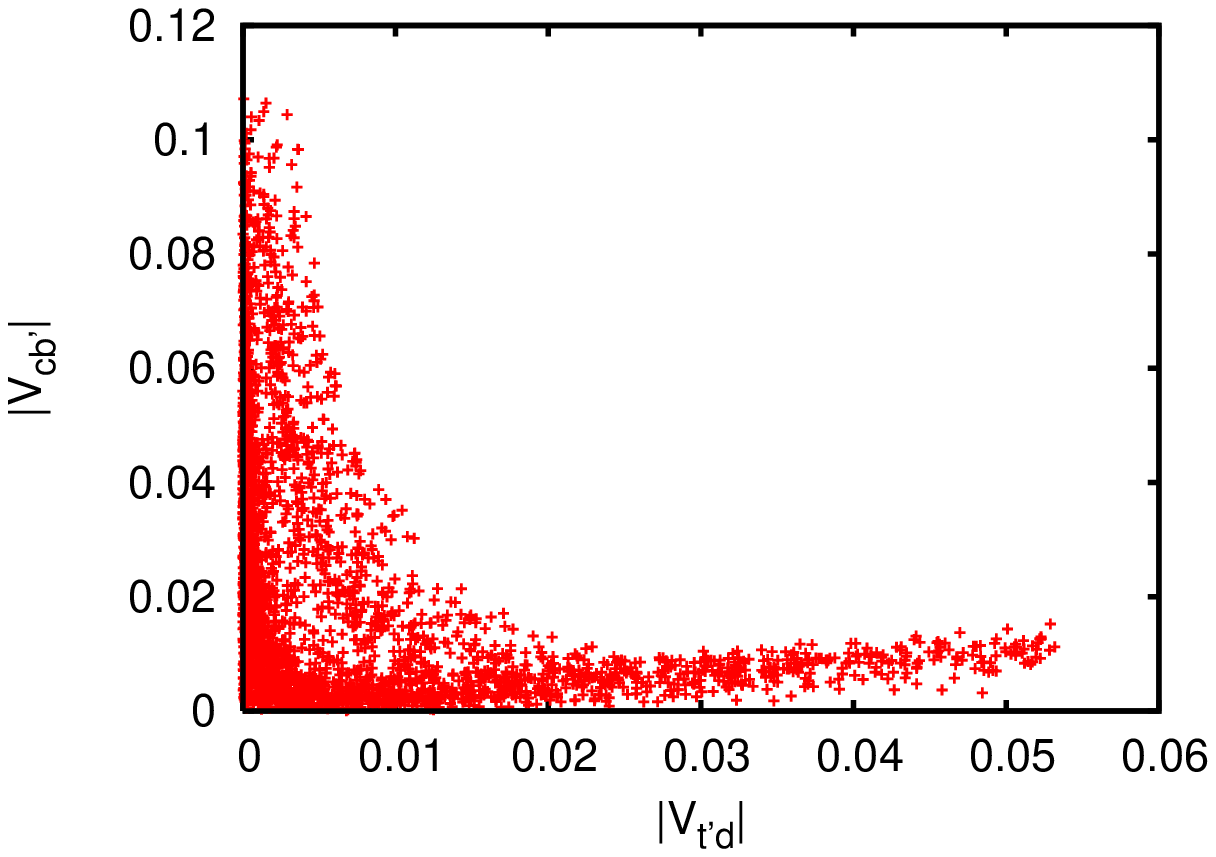}\hspace{.03\textwidth}
\includegraphics[width=.48\textwidth]{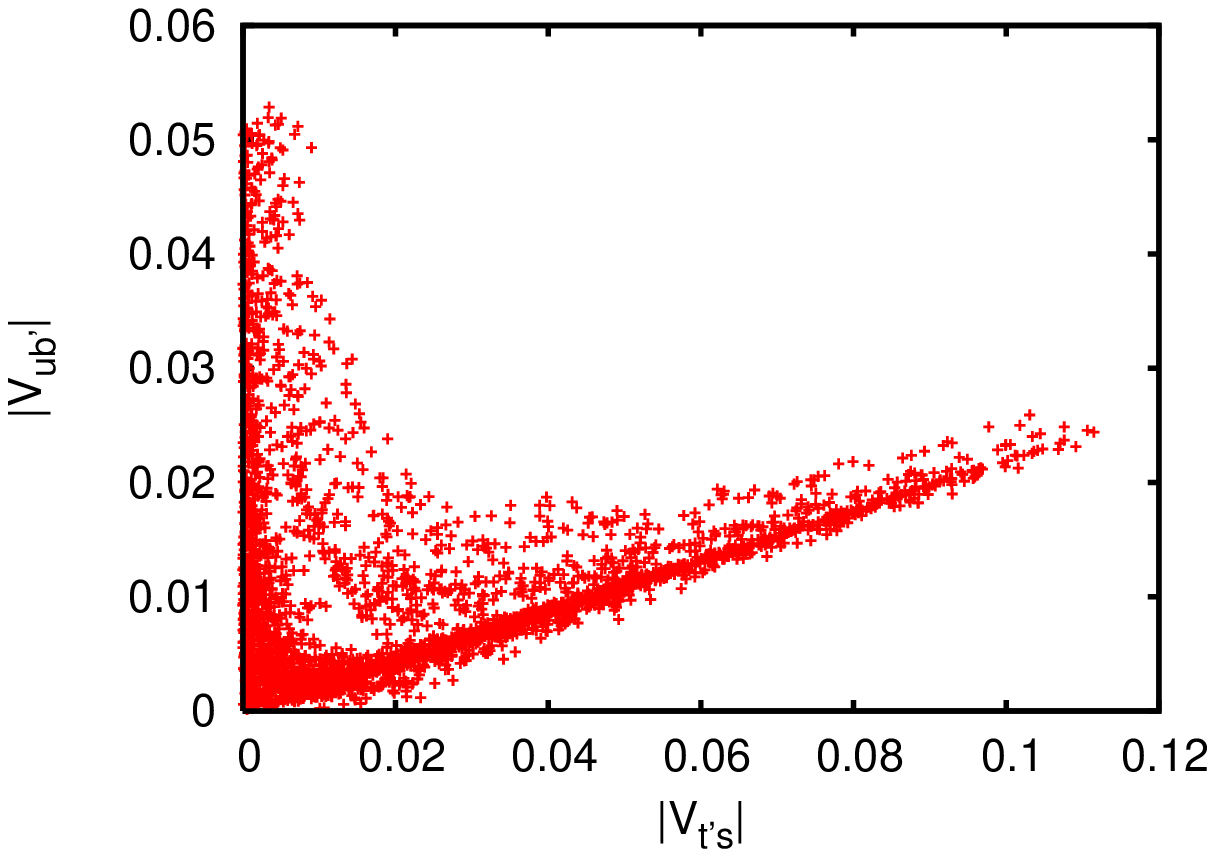}\\
\includegraphics[width=.48\textwidth]{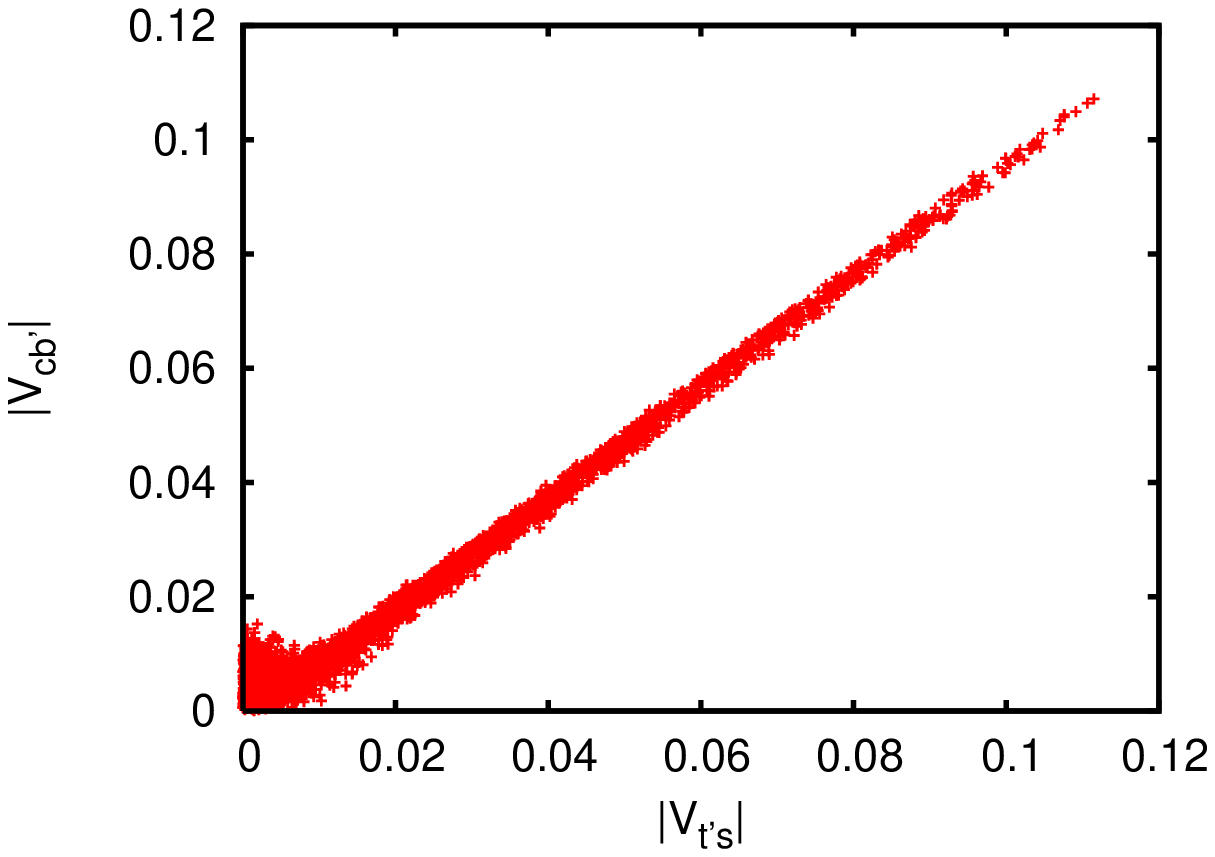}\hspace{.03\textwidth}
\includegraphics[width=.48\textwidth]{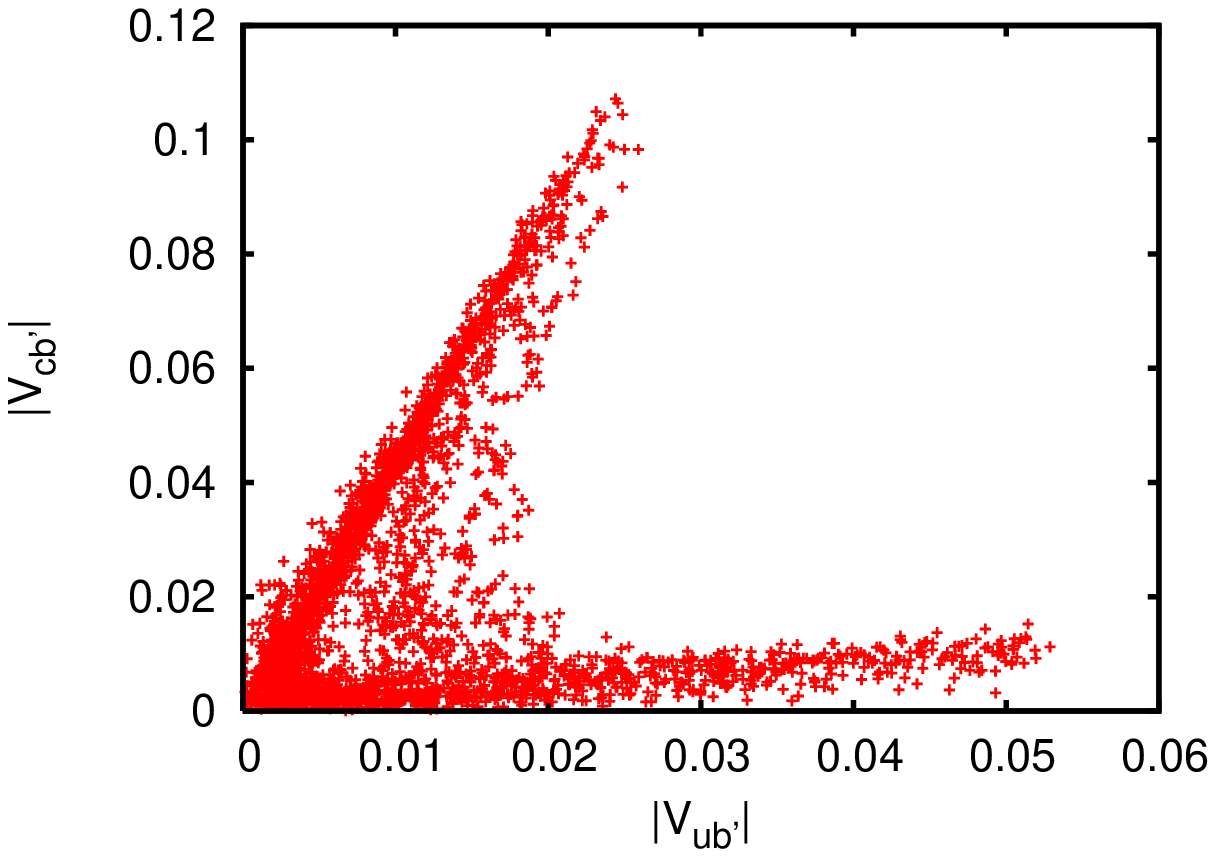}
\caption{Correlations between different new CKM4 elements.}
\label{fig2}
\end{figure}

In Fig. \ref{fig2} (upper-left panel) we show the correlation between $P$ and $Q$, larger 
values of $P$ corresponds to lower value of $Q$ and vice versa. We obtain such a correlation 
mainly due to the constraints from $\epsilon_K$ and $Br(K^+\to \pi^+\nu\bar\nu)$, although the upper bound 
on $P$ and $Q$ is coming from the other $B_d$ and $B_s$ data ( see Table \ref{tab1}). The expressions for 
$\epsilon_K$ and $Br(K^+\to \pi^+\nu\bar\nu)$ are sensitive to $\la^{t'}_{ds}$ i.e using these inputs 
we will get direct constraint on $\la^{t'}_{ds}$; as indicated in eq. \ref{paradep} $\la^{t'}_{ds}$ 
is proportional to the product of $P$ and $Q$. Therefore we will get direct constraint on the product 
not on individual $P$ or $Q$ and this is the reason why they follow the correlations shown. 
Similar correlation is possible between $V_{t'd}$ and $V_{t's}$ since they are proportional to $P$
and $Q$ respectively. We also show the correlations between some other CKM4 elements \footnote{Flavour data allows us
to get direct constraints on various products of CKM4 elements. The bounds on individual CKM4 element
are obtained using the constraints on the product couplings. Data for non-decoupling oblique corrections helps
to get tighter constraint on $|V_{t'b}|$ which helps to constraint  $|V_{t's/d}|$ from the bound on $|\la^{t'}_{s/d b}|$.}
; the plot of $|V_{t's}|$ as a function
of $|V_{cb'}|$ (lower-left panel) shows that $|V_{t's}| \approx |V_{cb'}|$ since leading order contribution to  
both the terms is proportional to $Q$. However, the plot of $|V_{t'd}|$ as a function of $|V_{ub'}|$ 
(upper-right panel) shows such a relationship only when $P >> Q$ since the leading order contribution to   
$|V_{t'd}|$ is proportional to $P$ whereas that for $|V_{ub'}|$ is proportional to a linear combination of 
$P$ and $Q$, see eq. \ref{eqn:v4g}. For relatively smaller values of $|V_{t'd}|$, of  ${\cal O}(10^{-3})$, $|V_{ub'}|$ 
could be as high as $0.02$; it is possible when $Q > P$ or alternatively when $|V_{t's}| \gsim 0.025$ 
(middle-right panel, Fig. \ref{fig2}). It also shows that larger values 
of $|V_{ub'}|$ are still possible which correspond to $P >> Q$ i.e for smaller values of $|V_{t's}|$. 
Similar feature can be observed in the correlation between $|V_{ub'}|$ and $|V_{cb'}|$ (lower-right panel). Middle-left 
panel of Fig. \ref{fig2} shows the correlation between $|V_{t'd}|$ and $|V_{cb'}|$ which is similar to the 
correlation between $P$ and $Q$ since leading order contribution in $|V_{cb'}|$ is $\propto Q$ and that for $|V_{cb'}|$ 
is $\propto P$.

\begin{figure}[t] 
\includegraphics[width=.48\textwidth]{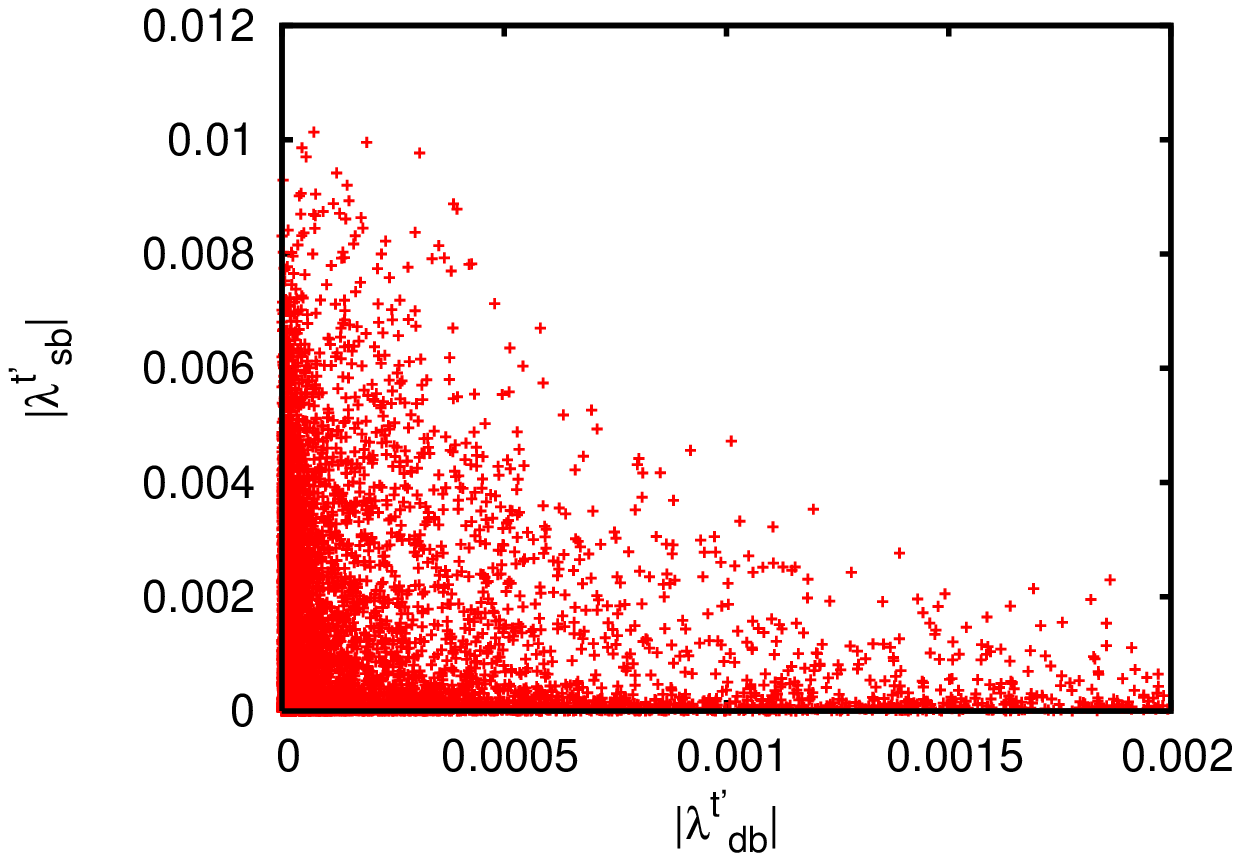}\hspace{.03\textwidth}
\includegraphics[width=.48\textwidth]{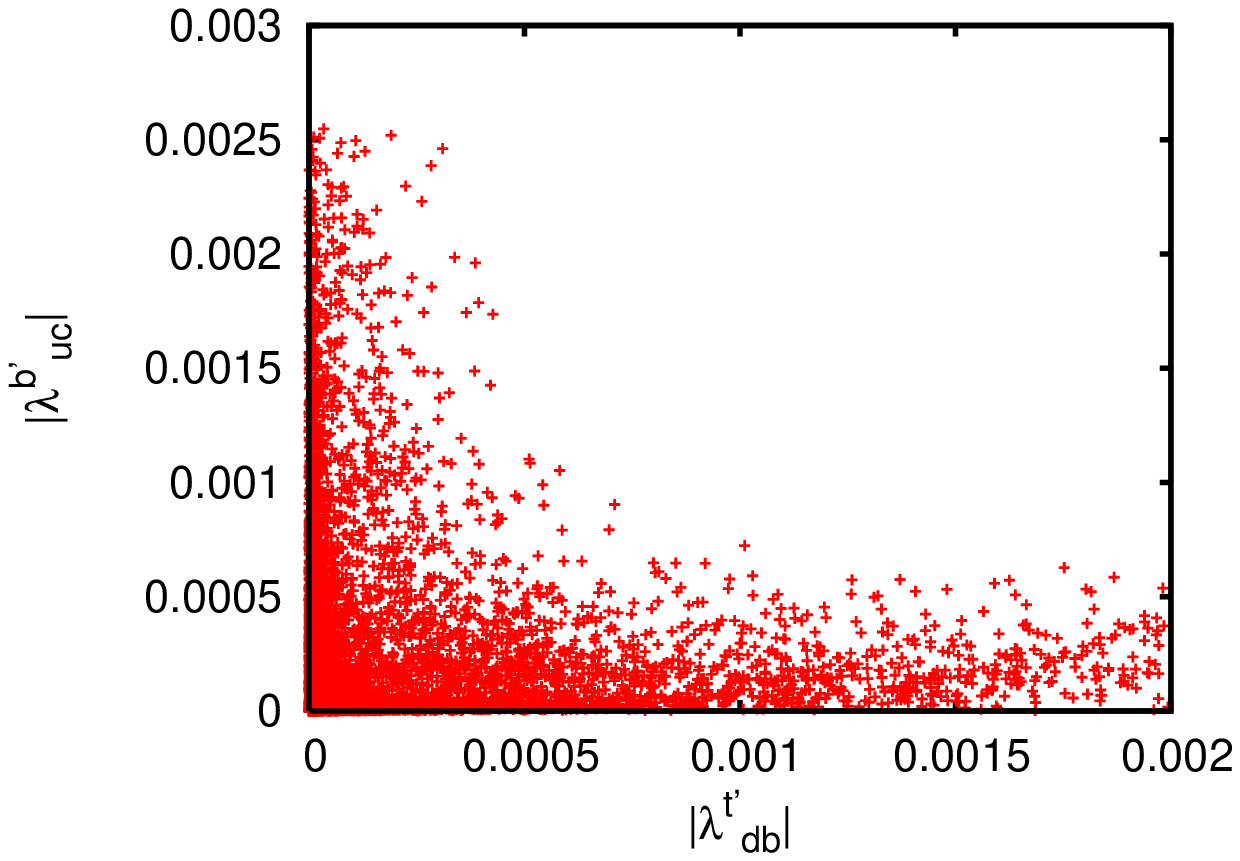}\\
\includegraphics[width=.48\textwidth]{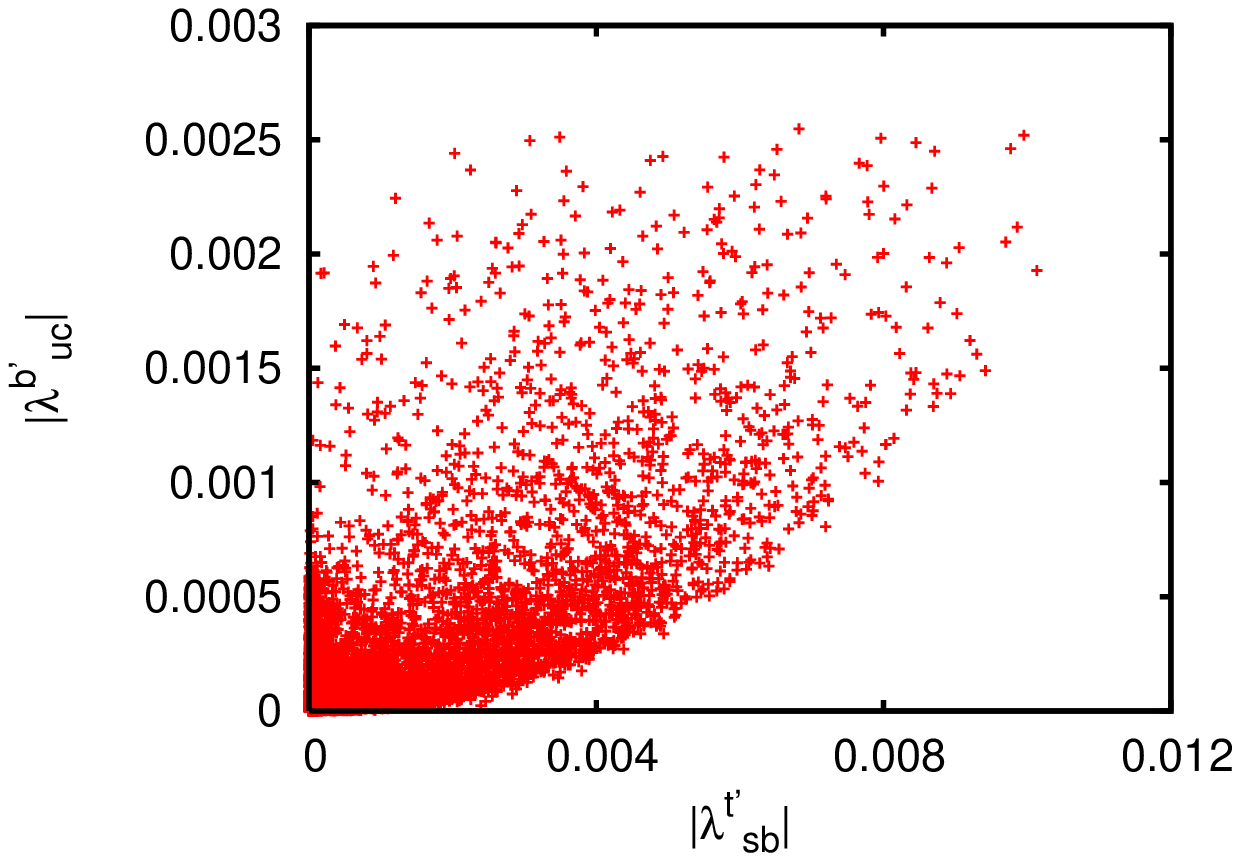}
\caption{Correlations between different CKM4 product couplings, $|\la^{t'}_{db}| = 
|V^{\ast}_{t'd} V_{t'b}|$ and $|\la^{t'}_{sb}| = |V^{\ast}_{t's} V_{t'b}|$ (upper left panel), 
$|\la^{t'}_{db}|$ and $|\la^{b'}_{uc}| = |V^{\ast}_{ub'} V_{cb'}|$ (upper right panel), 
$|\la^{t'}_{sb}|$ and $|\la^{b'}_{uc}|$ (lower panel). }
\label{fig3}
\end{figure}

The mathematical expressions for the product of CKM4 elements $|\la^{t'}_{db}|$ and $|\la^{t'}_{sb}|$ are given by
\begin{align}
|\la^{t'}_{db}| &= |V^{\ast}_{t'd} V_{t'b}| =  P r \la^4 \nonumber \\
|\la^{t'}_{sb}| &= |V^{\ast}_{t's} V_{t'b}| =  Q r \la^3,
\label{eqdsb}
\end{align}
whereas that for $|\la^{b'}_{uc}|$ can be obtained from eq. \ref{paradep} by taking its modulus, and we 
see that when $P<< Q$ it is $\approx Q^2 \la^5$.
In Fig. \ref{fig3} we show the correlations between the products of CKM4 elements; 
upper-left panel shows the correlation between $|\la^{t'}_{db}|$ and 
$|\la^{t'}_{sb}|$ which is similar to the correlation between $P$ and $Q$ (upper-left panel Fig. \ref{fig2}), as 
expected since the slope of the curve is given by $\frac{Q}{P \la}$. In the upper-right panel of Fig. \ref{fig3} 
we show the correlation between $|\la^{t'}_{db}|$ and $|\la^{b'}_{uc}|$ and note that $|\la^{b'}_{uc}|$ 
could be as large as $0.0025$ when $|\la^{t'}_{db}|$ is very small (say $< 0.0005$) i.e when $P << Q$ and vice versa.      
The most interesting one is the correlation between $|\la^{t'}_{sb}|$ and $|\la^{b'}_{uc}|$ (lower-panel Fig. \ref{fig3}); 
it shows an almost linear relationship between them which is prominent for larger values of $|\la^{t'}_{sb}|$ i.e for larger values 
of $Q$ due to strong $Q^2$ dependence of $|\la^{b'}_{uc}|$. It plays an important 
role in understanding the correlations between the {\it CP} asymmetries in $B_s$ and $D$ system; later we will discuss 
it in detail. The final remark from these discussions is that the allowed parameter space for the new CKM4 parameter 
space are highly correlated; random choices of the CKM4 parameters are not allowed, while doing so one 
has to be careful and the chosen values should be consistent with the appropriate correlations.

\begin{figure}[htbp]
\includegraphics[width=.48\textwidth]{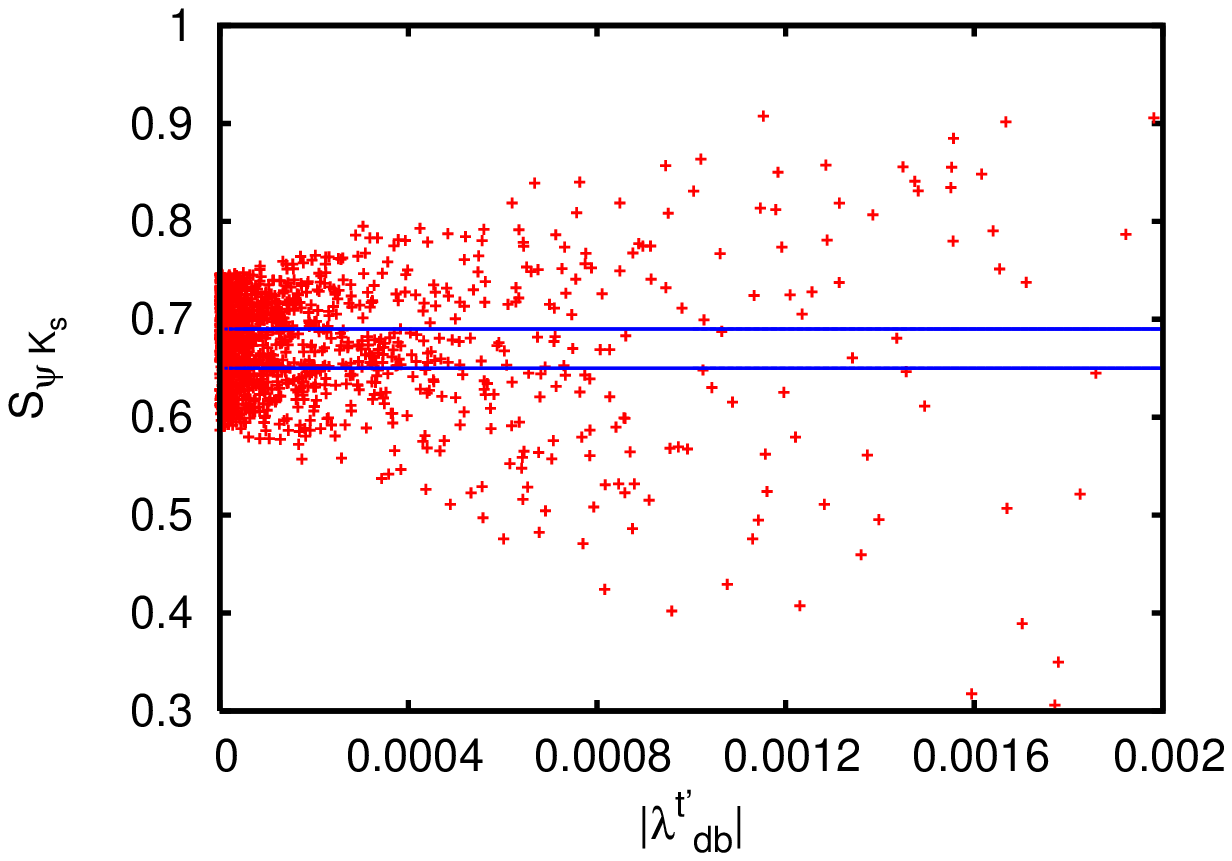}\hspace{.03\textwidth}
\includegraphics[width=.48\textwidth]{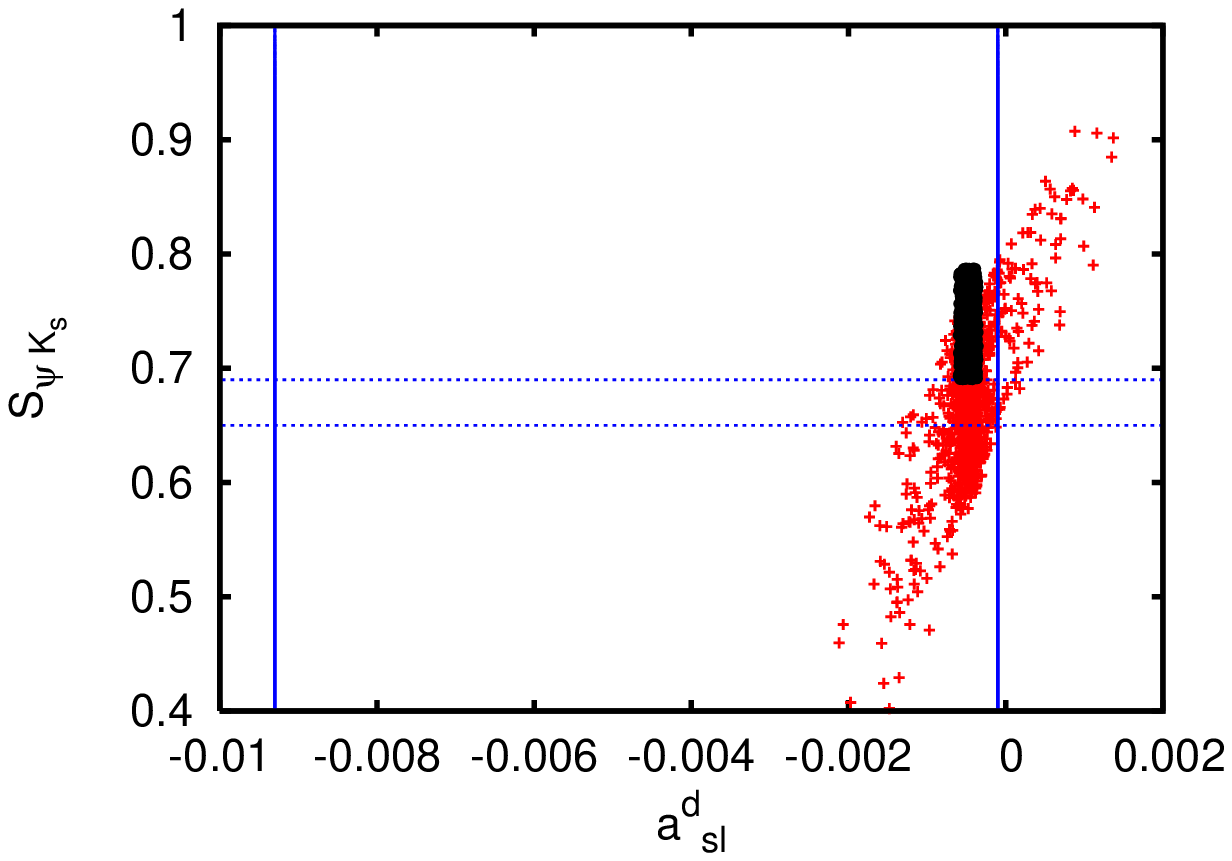}\\
\includegraphics[width=.48\textwidth]{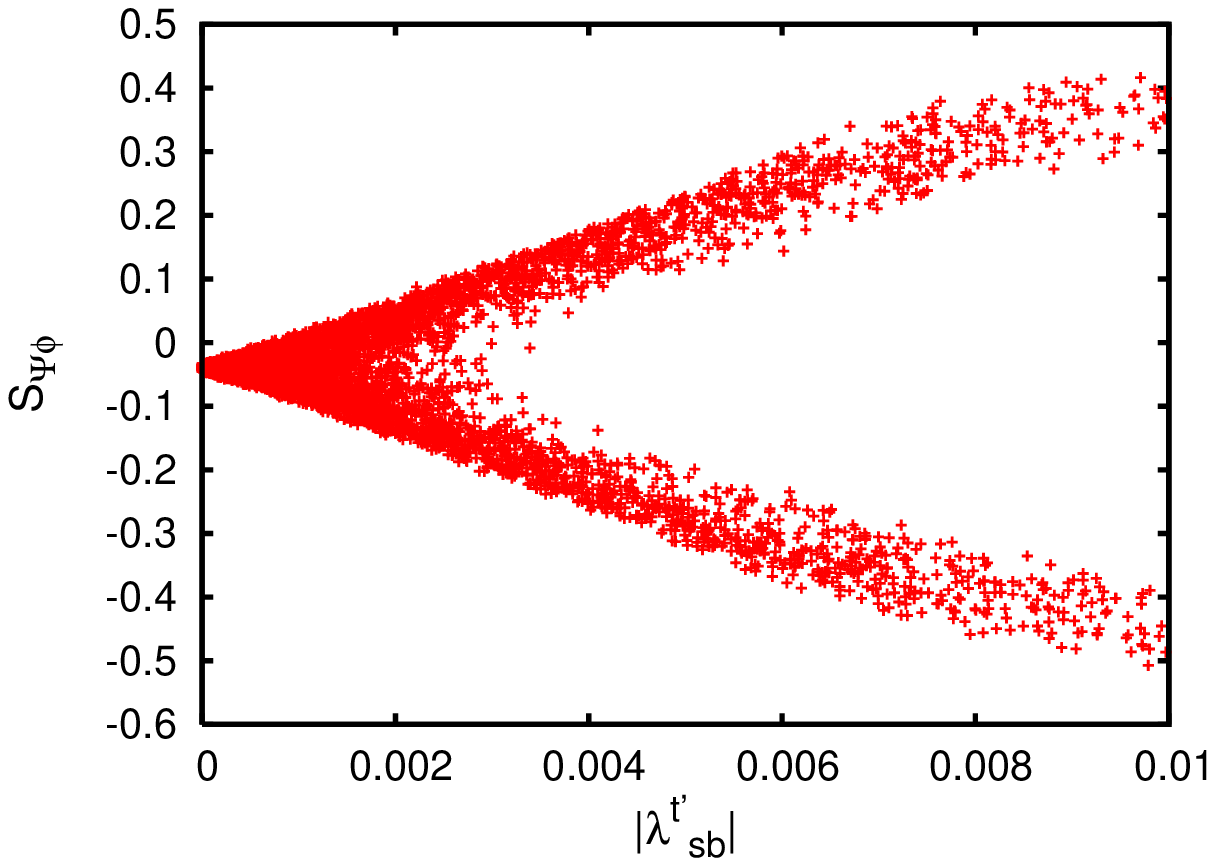}\hspace{.03\textwidth}
\includegraphics[width=.48\textwidth]{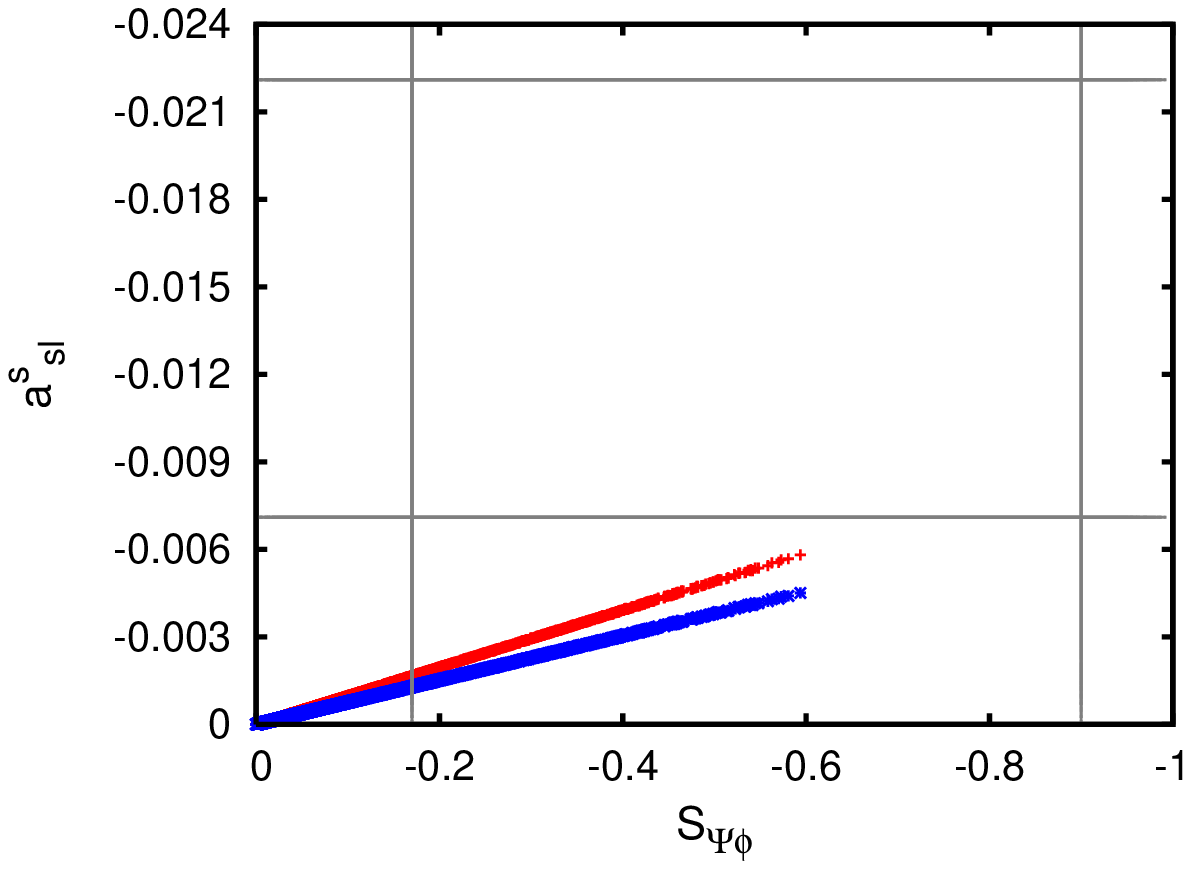}
\caption{Various correlations in SM4 are shown. Variation of CP asymmetries with the magnitude of the product couplings; 
$S_{\psi K_s}$ as a function of $|\la^{t'}_{db}|$ (upper-left panel), $S_{\psi\phi}$ as 
a function of $|\la^{t'}_{sb}|$ (lower-left panel). Correlations between the 
time dependent mixing induced CP asymmetries and semileptonic CP asymmetries for $B_d$ and 
$B_s$ are shown in upper and lower right panel respectively. Blue horizontal and vertical bands 
are the corresponding experimental ranges. In the upper-right panel SM allowed band (thick dark) is shown in 
$S_{\psi K_s}$ vs $a_{sl}^d$ plane.  
In the lower-right panel the blue and red regions 
correspond to $\Delta\Gamma_s$ with the uncertainties taken at $1\sigma$ and $2\sigma$ respectively 
and grey horizontal band corresponds to the experimental range for $a^s_{sl}$ and the vertical 
band is that for CP asymmetry.}
\label{fig4}
\end{figure}

Let us move to next part of our discussion where we show the effect of the fourth generation on different 
observables related to $B_d$, $B_s$ and $D$ system. In Fig. \ref{fig4} (upper-left panel) we plot CP 
asymmetry $S_{\psi K_s}$ as a function of $\la^{t'}_{db}$ and note that $S_{\psi K_s}$ 
can go down to $\approx 0.4$ or can reach around $0.9$ for large values of the product coupling $|\la^{t'}_{db}|$; 
so appreciable deviation from the present experimental measurement is, in principle, possible. We do not get any noticeable 
correlation between $S_{\psi K_s}$ with the phase $\dtpd$ of $\la^{t'}_{db}$. 
In the upper-right panel of Fig. \ref{fig4} we show the semileptonic asymmetry
$a^d_{sl}$ ($B_d$ system) as a function of $S_{\psi K_s}$. 
In SM4 the present experimental bound on $S_{\psi K_s}$ allows $a^d_{sl} \gsim -0.001$, whereas SM has a bound, $(-4.8 ^{+1.0}_{-1.2})\times 10^{-4}$ as shown by the black band in the Fig. \ref{fig4} (upper-right panel).

In the lower-left panel of Fig. \ref{fig4} we are showing the allowed regions for the CP asymmetry
$S_{\psi\phi}$ in $B_s\to \psi\phi$ as a function of $|\la^{t'}_{sb}|$, for
$ 375 \, {\it GeV} < m_{t'} < 575 \,{\it GeV}$, $S_{\psi\phi}$ is bounded by $ - 0.50 \lsim S_{\psi\phi} \lsim 0.50$,
the explicit dependence on $m_{t'}$ has been shown in our earlier papers \cite{SAGMN08,SAGMN10}.
It is also interesting to note that its magnitude increases with $|\la^{t'}_{sb}|$;
precise measurements of $S_{\psi\phi}$ will be helpful to put tighter constraints on $|\la^{t'}_{sb}|$ and 
the corresponding phase. Recently CDF and  DO  have updated their measurement of the CP-violating phase with data
sample corresponding to an integrated luminosity of $5.2\,fb^{-1}$ and $6.1\,fb^{-1}$
respectively.
The allowed 68\% C.L ranges are \cite{cdfnew,d0new}
\begin{align}
\phi_s^{\psi\phi} \,&\in \, \lt[ -0.04,-1.04 \rt]\cup\lt[-2.16,-3.10\rt], &  CDF\nonumber \\
                   & \in - 0.76^{+0.38}_{-0.36}\, (stat) \pm 0.02 (syst)  &   DO.
\label{cdfphi}
\end{align}
The corresponding $1\sigma$ ranges for $S_{\psi\phi} = \sin{\phi_s^{\psi\phi}}$ are given in Table \ref{tab:cp}.


In lower-right panel Fig. \ref{fig4}  we show the correlation \footnote{The plot corresponds            
to negative solution for $S_{\psi\phi}$, we do not show the points corresponding to the positive solution of $S_{\psi\phi}$
for which one should get a region symmetric to that shown in the figure.} between $S_{\psi\phi}$ and $a^s_{sl}$ (eq. \ref{aslq}) with \cite{uli_lenz} 
\bea
\Delta\Gamma_s^{SM} = 0.096 \pm 0.036  \hskip 20pt 68\% \,\,C.L\,\, ,
\label{dgsm}
\eea
taken at $1\sigma$ (blue) and $2\sigma$ (red). We note that its magnitude increases
with $S_{\psi\phi}$ as well as with $\Delta\Gamma_s$, as expected from eq. \ref{aslq},  the maximum 
allowed ranges are given by
\begin{align}
a^s_{sl} &\gsim - 0.004, \,\,\,\,\,\,\,\, \Delta\Gamma_s^{SM}\,@1\sigma,\nonumber \\
         &\gsim - 0.006, \,\,\,\,\,\,\,\, \Delta\Gamma_s^{SM}\,@2\sigma.
\end{align}

In Table \ref{tab:cp} we summarise the allowed ranges
for different CP observables in SM4, it includes time dependent CP asymmetries in
$B_d \to \psi K_s$, $B_s \to \psi\phi$ as well as the semileptonic asymmetries associated with $B_d$ and $B_s$
system (eq. \ref{aslq}). We also mention the corresponding experimental ranges and SM3 predictions
obtained with the inputs given in Table \ref{tab1}.

\begin{table}[htbp]
\begin{center}
\begin{tabular}{|l|l||l|l|}
\hline
CP observable & SM3 & Exp & SM4 ranges \\
\hline\hline
$S_{\psi K_s} = \sin 2\beta_d$ & 0.739 $\pm$ 0.049 & 0.67 $\pm$ 0.02 & 0.40 $\to$ 0.90 \\
\hline
$S_{\psi\phi} = \sin\phi^{\psi\phi}_s$ & -0.04 $\pm$ 0.002 & [-0.04,-0.86]\,\, CDF &  - 0.50 $\to$ 0.50 \\
 &     &  [-0.37,-0.90]\,\,  DO & \\
\hline
$a^d_{sl}$ & $(-4.8 ^{+1.0}_{-1.2})\times 10^{-4}$ &  -0.0047 $\pm$ 0.0046 & $>$ -0.002 \\
\hline
$a^s_{sl}$ & $(2.1\pm 0.6)\times 10^{-5}$ & - 0.0146 $\pm$ 0.0075 &  -0.006 $\to$ 0.006 \\
\hline
$\eta_f S_{CP}(D)$ & $\approx  -2 \cdot 10^{-6}$ & $-0.248\pm 0.496$\% & $-0.01 \to 0.01$ \\
\hline
$a_{sl}(D)$ & $\approx  1 \cdot 10^{-4}$ & & $-0.6 \to 0.6$\\
\hline 
\end{tabular}
\caption{Allowed ranges of different CP observables related to $B_d$, $B_s$ and $D^0$ systems in SM3 and SM4; current 
experimental status is also given.} \label{tab:cp}
\end{center}
\end{table}

\begin{figure}[t]
\includegraphics[width=.68\textwidth]{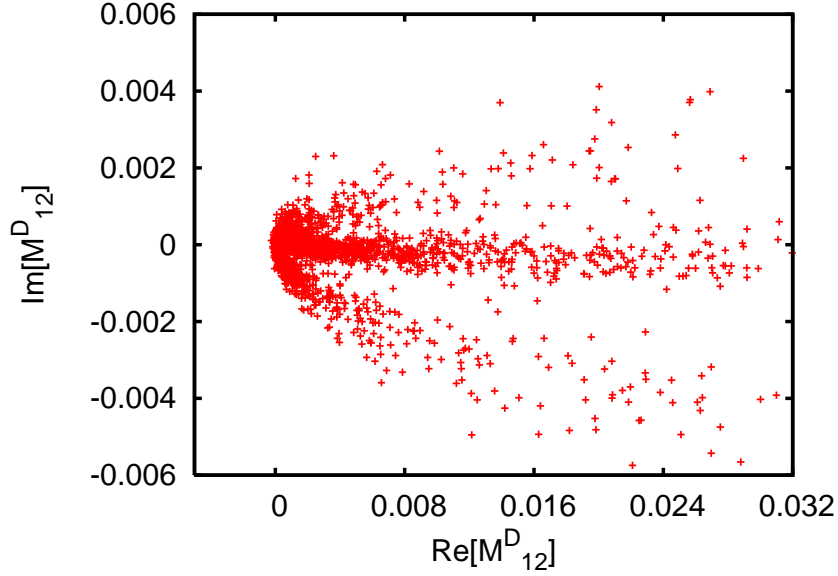}
\caption{Correlation between the real and imaginary part of the SD contribution to $M^D_{12}$.}
\label{fig5}
\end{figure}

In Fig. \ref{fig5} we show the correlation between the real and 
imaginary part of the short distance contribution to $\ddbar$ mixing.
Note that the magnitude of $Im(M^D_{12})$ could be as high as 0.6\%, which could 
be negative or positive; very small number of points are allowed for 
$Re(M^D_{12}) < 0$, however, it could be as high as $0.032$. These findings are in
good agreement with Ref. \cite{buras_charm}.

\begin{figure}[htbp]
\includegraphics[width=.48\textwidth]{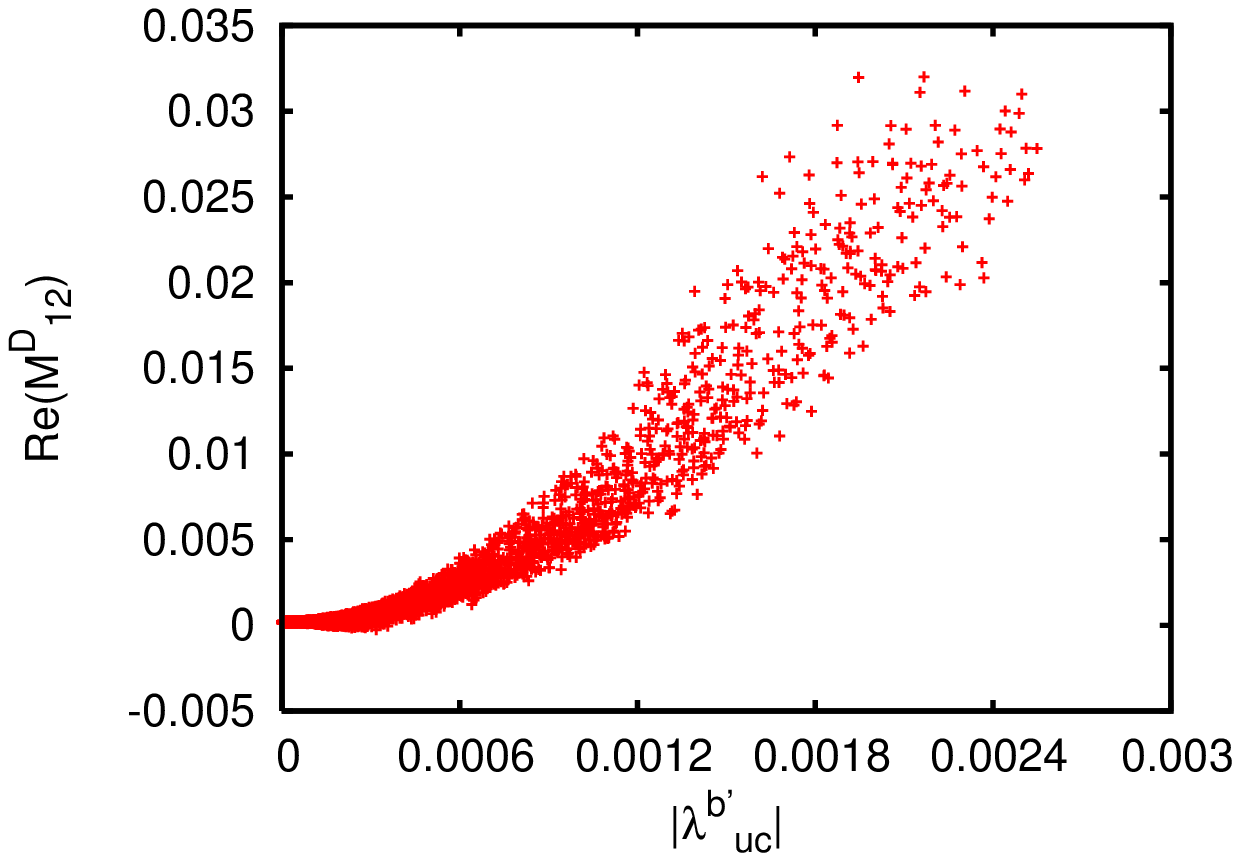}\hspace{.03\textwidth}
\includegraphics[width=.48\textwidth]{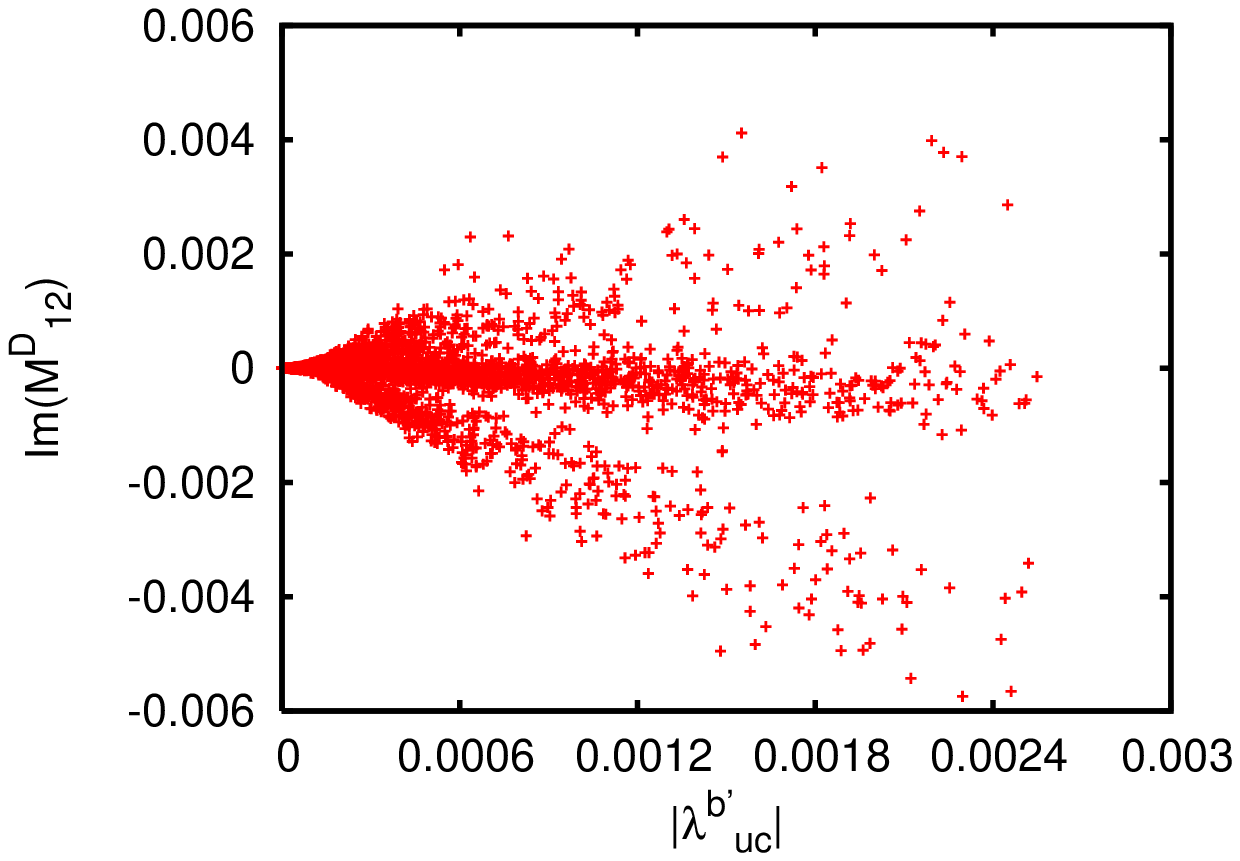}
\caption{Real (left panel) and imaginary (right panel) part of ${M^D_{12}}^{SD}$ as 
a function $|\la^{b'}_{uc}|$.}
\label{fig6}
\end{figure}

In Fig. \ref{fig6} we plot real (left panel) and imaginary (right panel) part of
$M^D_{12}$ as a function of $|\la^{b'}_{uc}|$ and note that in both the cases its
magnitude increases with the product coupling. In the case of the real part almost
all the allowed points are for $Re(M^D_{12}) > 0$, however, in case of imaginary
part we have both positive and negative solutions. As we noticed before (Fig. \ref{fig3}),
$|\la^{t'}_{sb}|$ has a linear relationship with $|\la^{b'}_{uc}|$; a tighter constraints 
on $|\la^{t'}_{sb}|$, which is possible to get by reducing the errors in the measurements of 
$B_d$ or $B_s$ observables, will be helpful to put tighter constrain on $\ddbar$ mixing.

\begin{figure}[htbp]
\includegraphics[width=.48\textwidth]{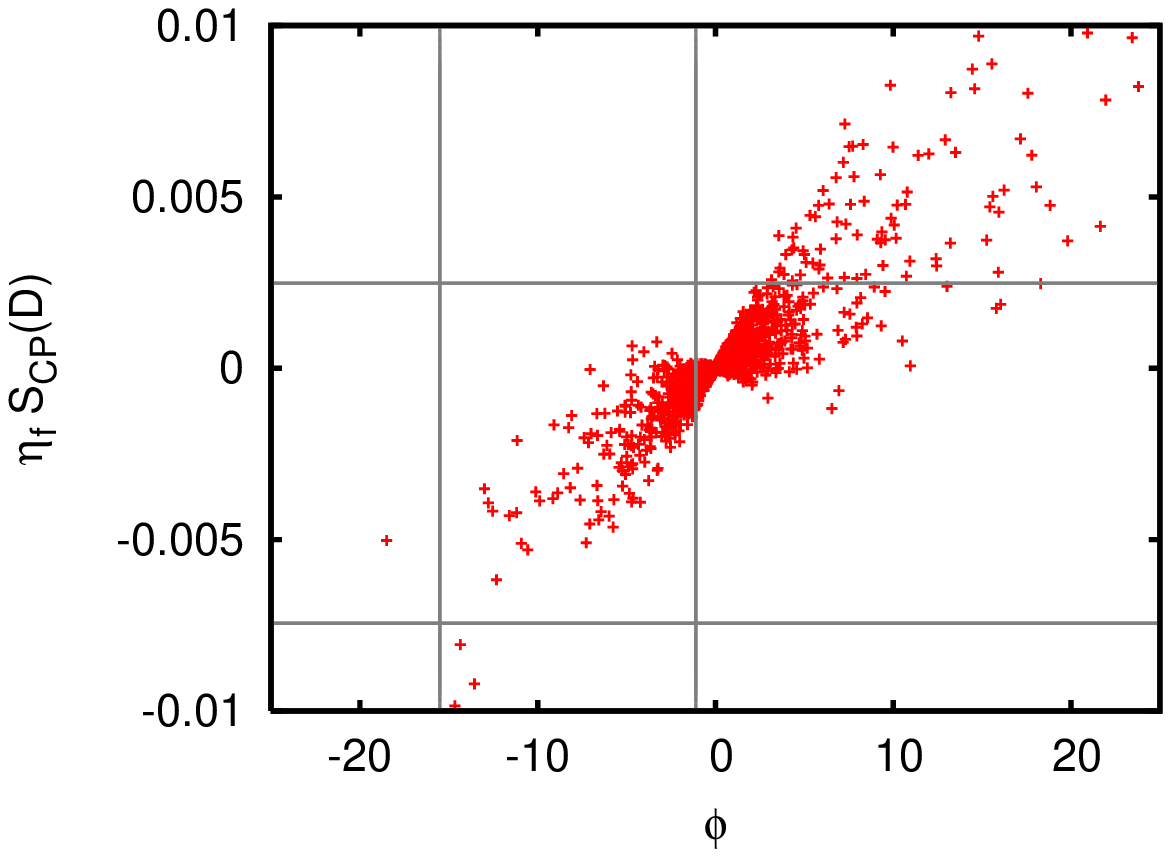}\hspace{.03\textwidth}
\includegraphics[width=.48\textwidth]{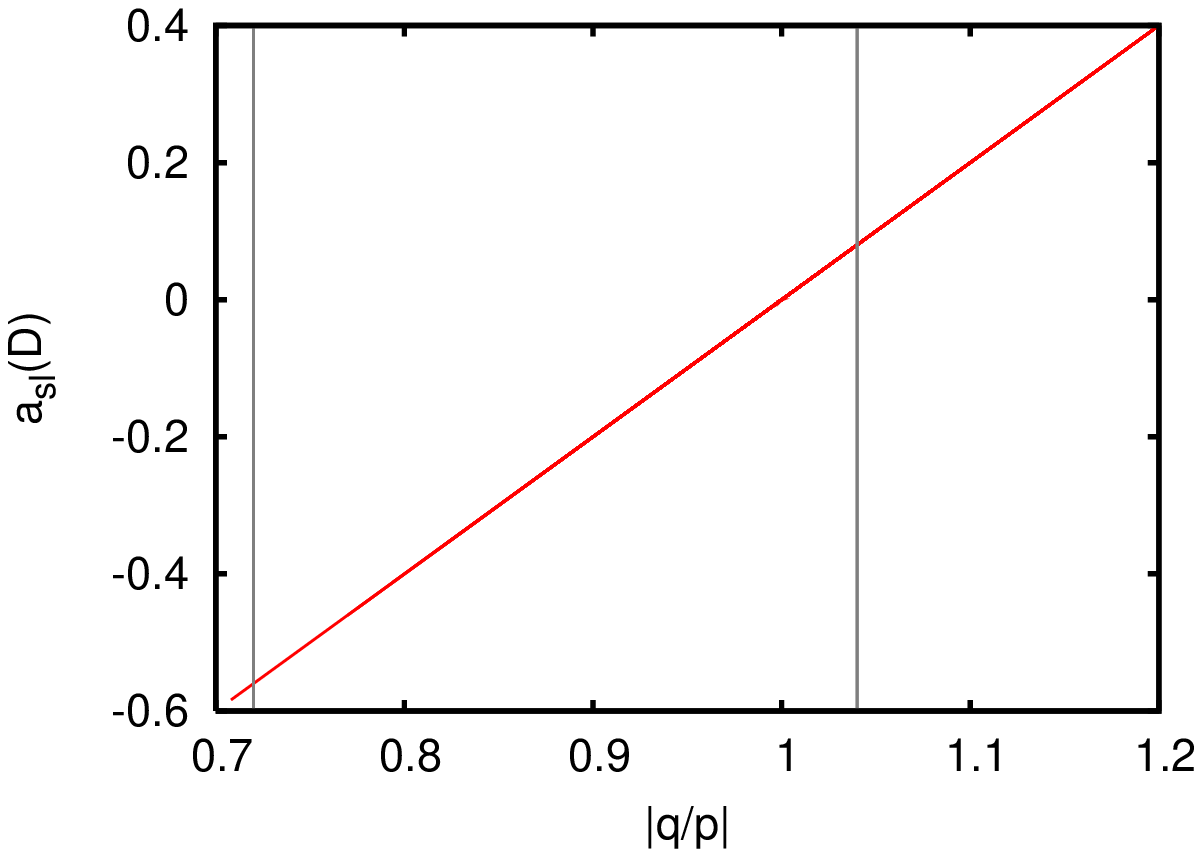}
\caption{Time dependent CP asymmetry for $D$ system, $\eta_f S_{CP}(D)$, as a function of 
the phase $\phi$ of $\frac{q}{p}$ (left panel), semileptonic CP asymmetry, $a_{sl}(D)$, 
as a function of $|\frac{q}{p}|$ (right panel) the corresponding SM value is 
${\cal O}(10^{-4})$. The grey horizontal and vertical bands represent the 
corresponding experimental ranges. }
\label{fig7}
\end{figure}

In Fig. \ref{fig7} we plot the time dependent CP asymmetry $\eta_f S_{CP}(D)$
(eq. \ref{eq:Sf}) and the semileptonic asymmetry $a_{sl}(D)$ (eq. \ref{eq:ASL}) in
the $D$ system as a function of the phase of $\frac{q}{p}$ (eq. \ref{eq:qp})  and
$|\frac{q}{p}|$ respectively; it could be directly compared with the correlations
shown in \cite{buras_charm}. We note that with the present experimental bound on the
phase of $\frac{q}{p}$ (eq. \ref{eq:expdata}), the magnitude of $\eta_f S_{CP}(D)$
could be enhanced up to the present experimental bound. On the other hand with the
present constraint on $|\frac{q}{p}|$, $a_{sl}(D)$ could be reduced to $-0.6$; again these 
results are also in agreement with Buras et. al \cite{buras_charm}.

\begin{figure}[htbp]
\includegraphics[width=.48\textwidth]{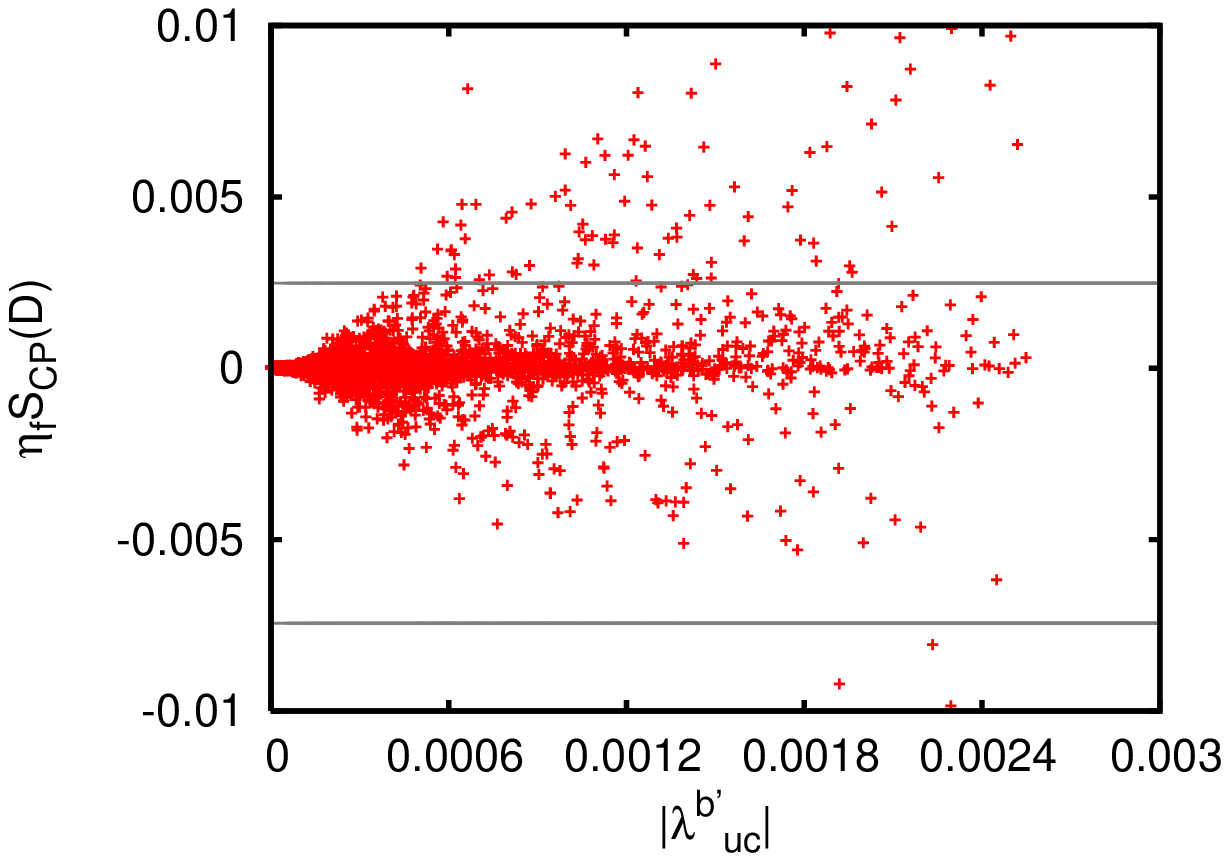}\hspace{.03\textwidth}
\includegraphics[width=.48\textwidth]{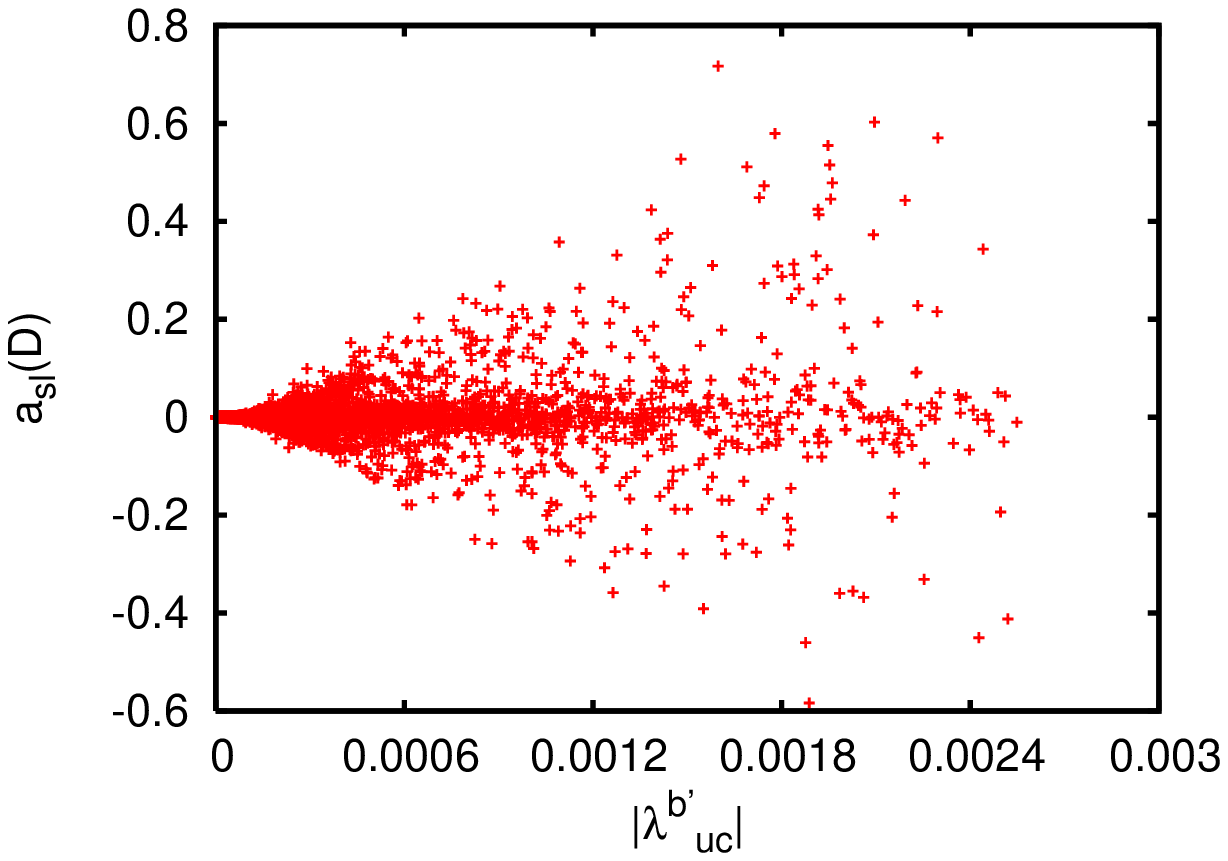}
\caption{$\eta_f S_{CP}(D)$ as a function of $|\la^{b'}_{uc}|$ (left panel), $a_{sl}(D)$ as a function
$|\la^{b'}_{uc}|$ (right panel).}
\label{fig8}
\end{figure}

In Fig. \ref{fig8} we plot $S_{CP}(D)$ and $a_{sl}(D)$ as a function of $|\la^{b'}_{uc}|$
and note that the magnitude of both may increase with $|\la^{b'}_{uc}|$; SM3 predictions and the 
allowed ranges in SM4 for the corresponding observables are summarised in Table \ref{tab:cp}. 
As discussed before (Fig. \ref{fig4}), 
the magnitude of $S_{\psi\phi}$ increases with the corresponding product
coupling, we also noticed that $|\la^{b'}_{uc}|$ increases with $|\la^{t'}_{sb}|$ which indicates a definite 
correlation between $S_{\psi\phi}$ and $\eta_f S_{CP}(D)$ \cite{buras_charm}. In the near future 
if we are able to put tighter constraints on $|\la^{t'}_{sb}|$, we will be able to get
strong limit on $\eta_f S_{CP}(D)$ and $a_{sl}(D)$ due to fourth generation effects.

\section{Conclusion}\label{concl}
This paper represents a continuation of our study of some of the
properties of SM4, Standard Model with four generations.
Herein we choose a specific representation for the 4X4 mixing matrix
and obtain constraints and correlations on its elements using available
data from K, B and D decays as well as electroweak precision tests and
oblique corrections and allowing  the $m_{t'}$ mass to
range from 375 to 575 GeV.  Constraints obtained are then used to
study the mixing induced and semi-leptonic CP asymmetries
in $B_d$, $B_s$ and in $D^0$. We find that SM4 allows S($B_d \to \psi
K_s$) to be closer to experiment thus alleviating a key
difficulty for SM3 that has been found in recent years.  SM4 allows
$a_{sl}^d$ to be bigger by a factor of O(3) . The
B-factories have a lot more data since they studied this asymmetry
some years ago \cite{hfag10}; it would be very worthwhile to update this bound.
On the other hand, $a_{sl}^s$ can be a lot bigger in SM4, and of
opposite sign,  than in SM3 where  it is essentially negligible. It
would also be very useful
to constrain this asymmetry as well as the linear combination ($A_{sl}^b$) \cite{d0dimuonprd}
of the two. Interestingly, the large same sign dimuon asymmetry recently
discovered by D0 \cite{d0dimuonprd} implies a rather large $a_{sl}^s$.  This has the
same sign as in SM4 though the central value of the D0 result is
somewhat larger than the expected range in SM4; however, the significance
of the D0 result is only about 2 $\sigma$ on $a_{sl}^s$.  These
asymmetries should be a high priority target for experiments at the
Tevatron as well as at LHCb.  In recent years Belle also has taken
appreciable  data at the $\Upsilon (5S)$ which should be used for
placing bounds on these asymmetries. In the future, these asymmetries should also be a very
useful target at the Super-B factories.\\

\vskip 20pt
Note Added: Very recently Ref~\cite{DL_10} presented constraints on SM4 using a
completely different representation of the 4X4 mixing matrix \cite{dgkm}.


\begin{acknowledgements}
SN thanks Theory Division of Saha Institute of Nuclear Physics (SINP) for 
Hospitality. SN's work is financially supported by NSERC of Canada.
The work of AS is suppported in part by the US DOE grant
\# DE-AC02-98CH10886(BNL).
\end{acknowledgements}

\end{document}